\documentclass[12pt]{book}
\usepackage{textpos}

\usepackage[utf8]{inputenc} 													
\usepackage[T1]{fontenc} 														

\usepackage{soul}							 										
\usepackage[portuguese, english]{babel} 										
\usepackage{cjhebrew}
\usepackage[space, extendedchars, multidot]{grffile} 							
\usepackage[official]{eurosym} 													

\usepackage{ragged2e} 															
\usepackage{setspace} 															
\usepackage{multicol} 															
\usepackage[flushmargin,bottom]{footmisc}									
\usepackage[all]{nowidow}														
\usepackage{courier} 															
\usepackage{tocloft}															
\usepackage{pdfpages}														
\usepackage{lastpage}														
\usepackage{textcomp}

\usepackage{amsfonts}															
\usepackage{amsthm}												  				
\usepackage{thmtools}												  				
\usepackage{mathtools}															
\usepackage{amsbsy} 															
\usepackage{amssymb}															
\usepackage{bbm}																
\usepackage{units}																
\usepackage{bbm}																
\usepackage{empheq} 															
\usepackage{centernot} 															
\usepackage{tensor}			 													
\usepackage{txfontsb}		 													
\usepackage{relsize,exscale}
\theoremstyle{definition}
\newtheorem{thm}{Theorem} 												
\newtheorem{prop}{Proposition} 											
\newtheorem{corol}{Corollary} 											
\newtheorem{lemma}{Lemma} 												
\newtheorem{dfn}{Definition}											

\DeclareMathOperator*{\argmax}{arg\,max}

\makeatletter
\let\save@mathaccent\mathaccent
\newcommand*\if@single[3]{%
  \setbox0\hbox{${\mathaccent"0362{#1}}^H$}%
  \setbox2\hbox{${\mathaccent"0362{\kern0pt#1}}^H$}%
  \ifdim\ht0=\ht2 #3\else #2\fi
  }
\newcommand*\rel@kern[1]{\kern#1\dimexpr\macc@kerna}
\newcommand*\widebar[1]{\@ifnextchar^{{\wide@bar{#1}{0}}}{\wide@bar{#1}{1}}}
\newcommand*\wide@bar[2]{\if@single{#1}{\wide@bar@{#1}{#2}{1}}{\wide@bar@{#1}{#2}{2}}}
\newcommand*\wide@bar@[3]{%
  \begingroup
  \def\mathaccent##1##2{%
    \let\mathaccent\save@mathaccent
    \if#32 \let\macc@nucleus\first@char \fi
    \setbox\z@\hbox{$\macc@style{\macc@nucleus}_{}$}%
    \setbox\tw@\hbox{$\macc@style{\macc@nucleus}{}_{}$}%
    \dimen@\wd\tw@
    \advance\dimen@-\wd\z@
    \divide\dimen@ 3
    \@tempdima\wd\tw@
    \advance\@tempdima-\scriptspace
    \divide\@tempdima 10
    \advance\dimen@-\@tempdima
    \ifdim\dimen@>\z@ \dimen@0pt\fi
    \rel@kern{0.6}\kern-\dimen@
    \if#31
      \overline{\rel@kern{-0.6}\kern\dimen@\macc@nucleus\rel@kern{0.4}\kern\dimen@}%
      \advance\dimen@0.4\dimexpr\macc@kerna
      \let\final@kern#2%
      \ifdim\dimen@<\z@ \let\final@kern1\fi
      \if\final@kern1 \kern-\dimen@\fi
    \else
      \overline{\rel@kern{-0.6}\kern\dimen@#1}%
    \fi
  }%
  \macc@depth\@ne
  \let\math@bgroup\@empty \let\math@egroup\macc@set@skewchar
  \mathsurround\z@ \frozen@everymath{\mathgroup\macc@group\relax}%
  \macc@set@skewchar\relax
  \let\mathaccentV\macc@nested@a
  \if#31
    \macc@nested@a\relax111{#1}%
  \else
    \def\gobble@till@marker##1\endmarker{}%
    \futurelet\first@char\gobble@till@marker#1\endmarker
    \ifcat\noexpand\first@char A\else
      \def\first@char{}%
    \fi
    \macc@nested@a\relax111{\first@char}%
  \fi
  \endgroup
}
\makeatother

\usepackage{graphicx}															
\usepackage{tabularx}															
\usepackage{ltablex}															
\usepackage{pdflscape}															
\usepackage{longtable} 															
\usepackage{tabu} 																
\usepackage{hhline} 															
\usepackage{booktabs} 															
\usepackage{multirow} 															
\usepackage[textfont=bf,hypcap]{caption}		 								
\usepackage{ragged2e}															
\usepackage{array}																
\usepackage{epstopdf}															
\usepackage{fancyref}															
\usepackage{fancyref}															
\usepackage{float}																

\usepackage{pgfplots}
\pgfplotsset{compat=1.17}
\usetikzlibrary{calc}
\usepackage{dcolumn}
\usepackage{subcaption}

\usepackage{enumitem}															
\newlist{todolist}{itemize}{2}
\setlist[todolist]{label=$\square$} 
\usepackage{pifont}
\usepackage{etoolbox}															
\usepackage{ifthen} 															
\usepackage{datetime} 															

\newdateformat{MonthDDYYYY}{%
  \monthname[\THEMONTH] \THEDAY, \THEYEAR
  }
\newdateformat{MonthYYYY}{%
  \monthname[\THEMONTH] \THEYEAR
  }

\usepackage[sf,sl,outermarks]{titlesec}
\usepackage{titleps}
\renewcommand*{\thesection}{\arabic{section}.}
\renewcommand*{\thesubsection}{\thesection\,\arabic{subsection}.}
\renewcommand*{\thesubsubsection}{\thesubsection\,\arabic{subsubsection}}

\titleclass{\part}{straight}
\titleclass{\chapter}{straight}

\titleformat{\part}[block]
  {\normalfont\Large\itshape\centering}{}{18pt}{\Large\bfseries}
\titlespacing*{\part}{0pt}{*1.5}{*0}											

\titleformat{\chapter}[block]
  {\normalfont\Large\bfseries\centering}{}{14pt}{\Large\bfseries}
\titlespacing*{\chapter}{0pt}{*1.5}{*2}											

\titleformat{\section}[block]													
{\normalfont\Large\bfseries}{\thesection}{12pt}{\Large\bfseries}		
\titlespacing*{\section}{0pt}{*1.5}{*0}											

\titleformat{\subsection}[block]
{\normalfont\large\bfseries}{\thesubsection}{12pt}{\large\bfseries}
\titlespacing*{\subsection}{0pt}{*1}{*0.8}

\titleformat{\subsubsection}[block]
{\normalfont\large\itshape}{\thesubsubsection}{11pt}{\large\itshape}
\titlespacing*{\subsubsection}{0pt}{*1}{*0.8}

\titleformat{\paragraph}[block]
{\normalfont\normalsize\bfseries}{\hspace{-1cm}}{11pt}{\bfseries\normalsize}
\titlespacing*{\paragraph}{0pt}{*0.5}{*0.2}

\usepackage{tocloft}

\usepackage[authoryear]{natbib} 												
\setcitestyle{authoryear,aysep={},open={(},close={)}}

\usepackage[hypertexnames=false,backref=page]{hyperref}															
\newcommand*{\hyref}[2]{\hyperref[#1]{#2 \ref{#1}}}

\hypersetup{
    pdffitwindow=false,									
    pdfstartview={XYZ null null 1.00},					
    pdfnewwindow=true, 									
    colorlinks=true,									
    linkcolor={red!50!black},							
    citecolor={blue!50!black},							
    urlcolor=black,										
	linkbordercolor={white},							
    pdfauthor = {Duarte Goncalves},
    pdfkeywords = {1, 2, 3},
    pdftitle = {},
    pdfsubject = {},
    pdfpagemode = UseNone,
    backref = true,
}
\bibpunct{(}{)}{;}{a}{,}{,} 													

\addto\extrasenglish{															

}

\makeatletter
\DeclareRobustCommand\citepos													
  {\begingroup\def\NAT@nmfmt##1{{\NAT@up##1's}}%
   \NAT@swafalse\let\NAT@ctype\z@\NAT@partrue
   \@ifstar{\NAT@fulltrue\NAT@citetp}{\NAT@fullfalse\NAT@citetp}}

\pretocmd{\NAT@citex}{%
  \let\NAT@hyper@\NAT@hyper@citex
  \def\NAT@postnote{#2}%
  \setcounter{NAT@total@cites}{0}%
  \setcounter{NAT@count@cites}{0}%
  \forcsvlist{\stepcounter{NAT@total@cites}\@gobble}{#3}}{}{}
\newcounter{NAT@total@cites}
\newcounter{NAT@count@cites}
\def\NAT@postnote{}

\def\NAT@hyper@citex#1{
  \stepcounter{NAT@count@cites}%
  \hyper@natlinkstart{\@citeb\@extra@b@citeb}#1%
  \ifnumequal{\value{NAT@count@cites}}{\value{NAT@total@cites}}
    {\if*\NAT@postnote*\else\NAT@cmt\NAT@postnote\global\def\NAT@postnote{}\fi}{}%
  \ifNAT@swa\else\if\relax\NAT@date\relax
  \else\NAT@@close\global\let\NAT@nm\@empty\fi\fi								
  \hyper@natlinkend}
\renewcommand\hyper@natlinkbreak[2]{#1}

\patchcmd{\NAT@citex}															
  {\ifNAT@swa\else\if*#2*\else\NAT@cmt#2\fi
   \if\relax\NAT@date\relax\else\NAT@@close\fi\fi}{}{}{}
\patchcmd{\NAT@citex}
  {\if\relax\NAT@date\relax\NAT@def@citea\else\NAT@def@citea@close\fi}
  {\if\relax\NAT@date\relax\NAT@def@citea\else\NAT@def@citea@space\fi}{}{}
\patchcmd{\NAT@cite}{\if*#3*}{\if*\NAT@postnote*}{}{}

\makeatother
\newcommand*{\headerleft}{}
\newcommand*{\headerright}{~}
\newcommand*{\footright}{\thepage}
\newcommand*{\footcenter}{~}
\newcommand*{\footleft}{~}

\newpagestyle{body}{
  \sethead[\ifthenelse{\value{page}=0}{}{\headerleft}][][\ifthenelse{\value{page}=0}{}{\headerright}]{\ifthenelse{\value{page}=0}{}{\headerleft}}{}{\ifthenelse{\value{page}=0}{}{\headerright}}
  \setfoot[\ifthenelse{\value{page}=0}{}{\footleft}][\ifthenelse{\value{page}=0}{}{\footcenter}][\ifthenelse{\value{page}=0}{}{\footright}]
  {\ifthenelse{\value{page}=0}{}{\footleft}}{\ifthenelse{\value{page}=0}{}{\footcenter}}{\ifthenelse{\value{page}=0}{}{\footright}}
}

\usepackage{usebib}
\setcitestyle{authoryear,aysep={},open={(},close={)}}
\newbibfield{journal}

\newtheoremstyle{named}{}{}{}{}{\bfseries}{.}{.5em}{\thmnote{#3}}
\theoremstyle{named}

\makeatletter
\declaretheoremstyle[
	spaceabove=15pt,
	spacebelow=15pt,
    headformat=\NAME\, \NUMBER: \NOTE,
    headpunct={},
    notefont=\bfseries,
	notebraces={}{},
	bodyfont=\normalfont,
	postheadspace=\newline,
	qed=\qedsymbol
]{mythmstyle}
\makeatother

\usepackage{amsthm}
\usepackage{afterpage}

\usepackage{footnote}
\makesavenoteenv{minipage}

\usepackage{etoc}

\titleformat{\section}[block]													
{\normalfont\Large\bfseries}{\thesection.}{14pt}{\Large\bfseries}		
\titlespacing*{\section}{0pt}{*1.3}{*0.2}											

\titleformat{\subsection}[block]
{\normalfont\large\bfseries}{\thesubsection.}{13pt}{\large\bfseries}
\titlespacing*{\subsection}{0pt}{*1}{*0}

\titleformat{\subsubsection}[block]
{\normalfont\bfseries}{\thesubsubsection.}{12pt}{\normalsize\bfseries}
\titlespacing*{\subsubsection}{0pt}{*1}{*0.8}

\renewcommand\thesection{\arabic{section}}
\renewcommand\thesubsection{\thesection.\arabic{subsection}}
\renewcommand\thesubsubsection{\thesubsection.\arabic{subsubsection}}
\usepackage{longtable}

\makeatletter
\newcommand*\TOCcompute@numwidths [2]{
	\begingroup
		\def\TOCnumwidthB {0pt}%
		\def\TOCnumwidthC {0pt}%
		\def\TOCnumwidthD {0pt}%
		\def\TOCnumwidthE {0pt}%
		\def\TOCnumwidthF {0pt}%
		\def\TOCnumwidthG {0pt}%
		\etocsetstyle{subsection}{}
			{\setbox0\hbox{\bfseries\etocthenumber\kern#2}}
			{\ifdim\wd0>\TOCnumwidthD\edef\TOCnumwidthD{\the\wd0}\fi}{}%
		\etocsetstyle{subsubsection}{}
			{\setbox0\hbox{\bfseries\etocthenumber\kern#2}}
			{\ifdim\wd0>\TOCnumwidthE\edef\TOCnumwidthE{\the\wd0}\fi}{}%
		\etocsetstyle{paragraph}{}
			{\setbox0\hbox{\bfseries\etocthenumber\kern#2}}
			{\ifdim\wd0>\TOCnumwidthF\edef\TOCnumwidthF{\the\wd0}\fi}{}%
		\etocsetstyle{subparagraph}{}
			{\setbox0\hbox{\bfseries\etocthenumber\kern#2}}
			{\ifdim\wd0>\TOCnumwidthG\edef\TOCnumwidthG{\the\wd0}\fi}{}%
		{\global\let\TOCnumwidthB\TOCnumwidthB
		\global\let\TOCnumwidthB\TOCnumwidthC
		\global\let\TOCnumwidthB\TOCnumwidthD
		\global\let\TOCnumwidthB\TOCnumwidthE
		\global\let\TOCnumwidthB\TOCnumwidthF
		\global\let\TOCnumwidthB\TOCnumwidthG}%
		\etocnopar
		\typeout{Next TOCs will use \TOCnumwidthB\space for chapter number width}%
		\typeout{Next TOCs will use \TOCnumwidthC\space for section number width}%
		\typeout{Next TOCs will use \TOCnumwidthD\space for subsection number width}%
		\typeout{Next TOCs will use \TOCnumwidthE\space for subsubsection number width}%
		\typeout{Next TOCs will use \TOCnumwidthF\space for paragraph number width}%
		\typeout{Next TOCs will use \TOCnumwidthG\space for subparagraph number width}%
		\endgroup
	}%
	\newcommand*\TOCcomputenumwidths [1][0.5em]{%
	\TOCcompute@numwidths {}{#1}%
	}%
	\newcommand*\TOCcomputelocalnumwidths [1][-10em]{%
	\TOCcompute@numwidths {local}{#1}%
}%
\makeatother

\makeatletter
\providecommand*{\hrefurl}{\hyper@normalise\hrefurl@}
\providecommand*{\hrefurl@}[2]{\hyper@linkurl{#2}{#1}}
\makeatother

\definecolor{mathematica1}{rgb}{0.368417, 0.506779, 0.709798}
\definecolor{mathematica2}{rgb}{0.880722, 0.611041, 0.142051}
\definecolor{mathematica3}{rgb}{0.560181, 0.691569, 0.194885}
\definecolor{mathematica4}{rgb}{0.922526, 0.385626, 0.209179}
\definecolor{mathematica5}{rgb}{0.528488, 0.470624, 0.701351}

\usepackage{dsfont,comment}

\usepackage{fouriernc}
\usepackage{charter}

\usepackage[top=1.5in, bottom=1.5in, left=1.5in, right=1.5in]{geometry}
\usepackage{libertine} 

\setlist[enumerate]{leftmargin=*,wide=0pt,itemsep=0pt,topsep=2pt}
\newcommand\blfootnote[1]{
  \begingroup
  \renewcommand\thefootnote{}\footnote{#1}
  \addtocounter{footnote}{-1}
  \endgroup
}

\title{The Dynamics of Social Instability}
\author{César Barilla (r) Duarte Gonçalves}
\AtBeginDocument{
  \setlength\abovedisplayskip{8pt}
  \setlength\belowdisplayskip{7pt}
  }
\begin{document}
\thispagestyle{empty}
\setcounter{page}{0}

\setcounter{footnote}{0}
\renewcommand{\thefootnote}{\fnsymbol{footnote}}
~\vspace*{-2cm}\\
\begin{center}\Large
    {\noindent
    \bfseries  The Dynamics of Instability}
\end{center}
\vspace*{1em}

\makebox[\textwidth][c]{
\begin{minipage}{1.2\linewidth}
\Large\centering
César Barilla\footnotemark
\;\textcircled{r}\;
Duarte Gonçalves\footnotemark
\end{minipage}
}
\setcounter{footnote}{1}\footnotetext{\setstretch{1}
Department of Economics, Columbia University;
cesar.barilla@columbia.edu. }
\setcounter{footnote}{2}\footnotetext{\setstretch{1}
Department of Economics, University College London;
duarte.goncalves@ucl.ac.uk. }

\blfootnote{
    We thank Jean-Paul Carvalho, Yeon-Koo Che, Joan Esteban, Teresa Esteban-Casanelles, Laura Doval, Navin Kartik, Elliot Lipnowski, Frederic Malherbe, Margaret Meyer, Christopher Sandmann, Ludvig Sinander and Yu Fu Wong for helpful comments and discussions. 
    We are also thankful for seminar participants at Oxford's `21 Southeast Theory Festival, SMYE `21, and SEA `22 Annual Conference, for useful feedback.
    Finally, we are grateful to the editor, Simon Board, and two anonymous referees for their suggestions.
}

\setcounter{footnote}{0}
\renewcommand{\thefootnote}{\arabic{footnote}}
\vspace*{-2em}

\begin{center} \large
March 10, 2023\\[30pt]
\end{center}

\begin{center}
\textbf{\Large Abstract}
\end{center}
\noindent\makebox[\textwidth][c]{
    \begin{minipage}{.85\textwidth}
        \noindent
        
        We study a model in which two players with opposing interests try to alter a status quo through instability-generating actions. 
        We show that instability can be used to secure longer-term durable changes, even if it is costly to generate and does not generate short-term gains. 
        In equilibrium, instability generated by a player decreases when the status quo favors them more. 
        Equilibrium always exhibits a region of stable states in which the status quo persists. 
        As players' threat power increases, this region shrinks, ultimately collapsing to a single stable state that is supported via a deterrence mechanism.
        There is long-run path-dependency and inequity: although instability eventually leads to a stable state, it typically selects the least favorable one for the initially disadvantaged player. 

        ~
    \end{minipage}
}
\newpage

Instability is an essential component of conflicts. Two political parties in a continuous competition for voters' support can stir unpredictable changes by leaking rumors or taking outrageous stances that will prompt controversy. Two countries fighting over disputed territories can adopt high-risk strategies that are often as likely to succeed as to backfire: instigating internal rebellions, launching propaganda campaigns or appealing to international organizations. Rebellious groups often seek to create agitation in the hope that they can profit from a volatile situation. Sometimes actions have no predictable effect other than to increase uncertainty, and instability itself becomes a means to an end: groups with diametrically opposed interests can generate instability strategically to advance their agenda.

In this paper, we study the strategic implications of using instability as an instrument in situations of conflict.
We consider a model in which two forward-looking players accrue bounded constant-sum gross flow payoffs. 
At any moment, players can pay a cost to increase the volatility of a process that determines the status quo division of payoffs. 
In particular, instability has a symmetric effect everywhere but at the extreme states, where it can only reflect the process towards less extreme states.

We show instability is an effective device in such situations. 
If the only thing a player can do is to destabilize the status quo in a way that change is equally likely to be favorable or unfavorable, instability cannot offer any advantage in the short-term. 
Furthermore, even if instability were to lead to a more favorable situation for the player, it could be met with additional instability by others with opposing views, further depressing the incentives to take action. 
But when a player has nothing to lose, instability seems like a natural instrument to oppose an excessively unfavorable status quo. 
We show how a lower bound on the negative consequences of creating instability provides option value that can be exploited by patient players, even when gains and losses are equally likely in the short run. 

We identify two key properties of players' optimal (Markovian) behavior. 
First, an optimal volatility strategy in response to any strategy of the opponent is characterized by a threshold mechanism: players continuously generate positive volatility at situations less favorable than a target ``satisficing'' state, and no instability at more favorable ones. 
Second, best responses to monotone strategies are monotone, creating more instability at less favorable states. 
Because gains over the status quo are driven by the option value conferred by the lower bound on how unfavorable the state can be, at a more favorable status quo, this option value decreases, and players become more conservative as they stand to lose more. 

We then prove existence and provide a complete constructive characterization of the set of equilibria.
An intuitive decoupling argument lies at the heart of this characterization: at most one player creates instability at any given moment. 
As instability yields no short-term gains on the status quo and players have diametrically opposed interests, they cannot both expect to benefit from it. 
As a result, equilibria are completely characterized by two thresholds, defining two regions of instability, and a stable region wherein the status quo prevails. 
Instability arises in the most extreme states, and the player who least favors the status quo creates instability to strive for change. 
Instability is used as a tool to push back against an extreme status quo, and more extreme states foster greater instability. 
In contrast, in the stable region --- corresponding to relatively more moderate states --- neither player sees advantage in destabilizing the status quo. 

While equilibrium stable states always exist, these can be either the expression of accommodating equilibrium behavior, or of a balance-of-power mechanism. 
In the former, players never push back against instability triggered by their opponent and so each player pursues gains on the status quo by generating instability exactly as if they faced no opposition. 
Such accommodating behavior occurs when impatience and costs to instability are high enough, which, owing to the threshold structure of best responses, supports a unique equilibrium.
This unique equilibrium generically features a continuum of stable states: those that are satisfactory to both players and, if perturbed, would not trigger any instability.
The situation fully reverses when players are patient and costs to instability are low enough: multiple equilibria arise, and each is characterized by a unique stable state. 
Further, this unique stable status quo emerges as resulting from players actively pushing back against their opponent's attempts to advance their prospects. 
Equilibrium behavior is then characterized as a balance-of-power mechanism at the stable status quo: the knowledge that the opponent will trigger social instability if the status quo is perturbed to their detriment deters the player from pursuing further improvements. 

Equilibria also exhibit clear monotone comparative statics. 
We find that lowering a player's costs to creating instability shifts the set of stable states in a strong set order sense toward states the player prefers, as the player is willing to generate more instability to pursue their goals.  

Finally, we discuss the dynamics of instability in our model.
We show that, regardless of the starting point, the process converges almost surely to an equilibrium stable state.
Nevertheless, we note a form of path dependency: if the process starts in a player's instability region, it will converge to that player's least favorable stable state.

Our paper contributes to the literature studying theoretical models of conflict.
Paraphrasing \citet[p. 387]{Fearon1995IO}, conflict is ``a gamble whose outcome may be determined by random or otherwise unforeseeable events.''
This observation motivated the modeling of conflict using contests, that is, situations in which players exert costly effort to affect their relative likelihood of obtaining a more favorable outcome.
Starting with the seminal work of \citet{Tullock1980Ch} in studying political party competition, several papers use this modeling device to study issues related to conflict and competition, including
conflict over the appropriation of rents \citep{BeskeyPersson2011QJE,Powell2013QJE}, 
lobbying \citep{BayeKovenockdeVries1993AER,CheGale1998AER}, 
territorial expansion \citep{BuenodeMesquita2020JoP,DziubinskiGoyalMinarsch2021JET}, and
how inequality affects the intensity of social conflict \citep{EstebanRay1999JET,EstebanRay2011JEEA}. 
Closer in spirit to this paper, \citet{FangNoe2016WP,FangNoe2022JLE} study risk-taking behavior in contest settings under a mean-performance constraint.
Our main contribution relative to the existing literature on conflict is to introduce a novel instability mechanism and relate it to key concepts and phenomena in the dynamics of conflict.
Our model gives qualitatively reasonable predictions for the dynamics of instability and, in doing so, highlights that instability need not be a purely exogenous byproduct, but rather a powerful and important instrument in situations of conflict.

Instability gives rise to two phenomena typically present in other models of conflict.
First, the fact that the disadvantaged player is the one who triggers instability is reminiscent of the idea that excessively unequal outcomes will trigger conflict \citep{Fearon1995IO} and that laggards choose more risky strategy in R\&D or sports \citep{Cabral2003JEMS,AndersonCabral2007RAND}, and, more broadly, consistent with the idea of ``gambling for resurrection'' \citep{KrakelNiekenPrzemeck2018EJPE,CalverasGanuzaHauk2004JRE}.
Second, although modeled in a different manner in either \citet{JacksonMorelli2009QJPS} or \citet{ChassangPadroiMiquel2010QJE}, the common theme of deterrence appears in our model in instances where a single state emerges as stable in equilibrium, and its stability is supported only by the fact that each of the two players with opposing interests would escalate conflict were the status quo affected.

Another related strand of the literature pertains to tug-of-war models and wars of attrition. 
In \citepos{MoscariniSmith2011WP} continuous-time analogue of the model by \citet{HarrisVickers1987REStud}, players with antagonistic preferences exert effort to increase the probability a state moves toward their preferred outcome, controlling the drift rather than the volatility of the process. 
\citet{AgastyaMcAffee2006WP} consider a related model with drift control and absorbing boundaries, in which stability obtains at intermediate states in draw equilibria because drift controls cancel each other.
\citet{GulPesendorfer2012REStud} and \citet{Gieczewski2020WP} consider war of attrition settings where players effectively control variance by choosing when to stop the payoff process.
Since in these models payoffs are accrued only when conflict stops and the extreme states are absorbing, close enough to the boundary, the winning side strives for a definitive victory and the losing side concedes, leading to stability; at intermediate states, there is conflict (hence instability) to determine to which side the scale will tip. 
This prediction reverses in our setting because there is no definitive victory.
Since the losing side has much less and the winning side more to lose, instability becomes a potent tool at extreme states, but too risky at intermediate ones.

Lastly, our paper contributes to a growing literature on games in continuous time.
Although the use of differential methods for zero-sum games dates back to the seminal work of \cite{Isaacs1965}, a number of recent contributions have effectively used stochastic calculus and differential equations techniques in continuous time games.\footnote{
    See \citet{Sannikov2007Ecta} for applications to repeated games, or \citeauthor{DaleyGreen2012Ecta} \citeyearpar{DaleyGreen2012Ecta,DaleyGreen2020AER} and \citet{Ortner2019GEB} for applications to bargaining with a continuous inflow of news and evolving bargaining power.
}
As other recent papers in economics \citep[e.g.][]{FaingoldSannikov2011Ecta,KaplanMollViolante2018AER,AchdouHanLasryLionsMoll2021REStud,Lester2020WP,KuvalekarLipnowski2019,EscudeSinanderWP} and a wealth of applications in finance,\footnote{
    See the monographs by \cite{FlemingSoner2006} or \cite{Pham2009Book} for more detail.
}  we rely on viscosity solutions to solve a non-smooth optimal control problem.
Building on \citet{Lions1986Proc}, this paper provides a technical contribution to this literature by proving existence and uniqueness of optimal control of volatility of a reflecting process under relaxed regularity conditions.
We hope that the present paper also serves to illustrate the usefulness of this approach for obtaining precise characterizations in economic applications while imposing minimal assumptions.

The remainder of the paper is organized as follows:
\hyref{sec:model}{Section} introduces the model. 
In \hyref{sec:controlproblem}{Section}, we give a detailed characterization of optimal instability strategies by studying the best response to a fixed opponent strategy; we pay particular attention the benchmark case when the opponent is inactive and a single player controls the volatility.
We use these results in \hyref{sec:equilibrium}{Section} to construct and characterize equilibria, and, in \hyref{sec:dynamics}{Section}, we discuss the equilibrium dynamics of the status quo: namely, convergence towards a stable state.
\hyref{sec:discussion}{Section} discusses some natural variations of our model.

\section{The Model}
\label{sec:model}
    We now introduce our model. 
    Time is continuous and indexed by $t \in \mathbb{R}_+$. 
    The state at time $t$ is given by $X_t \in [0,1]$, corresponding to a status quo; players $A$ and $B$ have opposing preferences over the status quo captured by constant-sum flow payoffs. 
    Player $A$ strictly prefers higher values of the status quo, whereas $B$ favors lower ones, and we remove any intrinsic incentive to generate instability by considering risk-neutral preferences. 
    Given these assumptions, it is without loss to normalize player $A$'s gross payoff at time $t$ to be given by $X_t$ and player $B$'s by $1-X_t$. 

    The state evolves randomly and continuously over time according to the following stochastic differential equation with reflection:
    \begin{align*}
        dX_t = \sqrt{2(\alpha_t+\beta_t)} dB_t - dK_t,
    \end{align*}
    \noindent where $B_t$ is a standard Brownian motion, $\alpha_t \geq 0$ and $\beta_t \geq 0$ are non-negative adapted processes controlled by players $A$ and $B$ respectively, and $K_t$ denotes the regulator process that reflects the process within $[0,1]$ when it hits either bound and is inactive in the interior --- i.e. if $X_t \in (0,1)$ we have $dK_t = 0$.\footnote{
        The presence of the regulator process $K_t$ is purely a technical device used to define a process whose infinitesimal variations essentially follows $dX_t = \sqrt{2(\alpha_t+\beta_t)} dB_t$ but where an inward push compensates every variation that would push the process outside of the bounded domain $[0,1]$; $K_t$ precisely defines this compensation to ensure that we have defined a process over $[0,1]$. 
        We give more technical details on the definition of the process in \hyref{sec:appendix:preliminaries}{}.
    } 
    
    This captures the idea that instability has a symmetric effect everywhere but at the boundary. 
    Over a small time interval, the change in the status quo is exclusively driven by instability: 
    at any instant, $X_t$ goes either up or down with equal probability, except at the boundaries ($0$ and $1$), where it simply cannot become more extreme. 
    Everywhere in the interior, the status quo changes in a purely noisy manner. 
    
    A key assumption is that no player can get a negative flow payoff --- intuitively, in sharing a finite resource, one cannot have less than ``nothing'' (nor more than ``everything''). 
    The reflecting boundaries express the fact that even when some player reaches the lower bound of their payoffs, the game does not terminate. 
    This contrasts with models with absorbing boundaries\footnote{
        As those discussed above, e.g. \citet{GulPesendorfer2012REStud,Gieczewski2020WP,MoscariniSmith2001Ecta,AgastyaMcAffee2006WP}.
    } where the game stops upon reaching an extreme point. 
    Those are more likely to be applicable to situations with a clear end-point (an election, a patent race, a sports match) and terminal payoff, whereas our model is more adequately describing situations of repeated interaction without a definite ending where payoffs continuously accrue (competition between political parties, long-lasting dispute over territories between countries, protracted wars and rebellions).
    Although continuity of the process is essential in capturing the desired intuition --- because over a small time interval the probability of hitting either bounds is zero, a form of local symmetry is ensured --- the fact that players control the level of instability means that the state can change extremely quickly, or not at all.

    Players $A$ and $B$ respectively control $\alpha_t$ and $\beta_t$ --- how much effort each puts into destabilizing the status quo. 
    Total instability effort $\alpha_t+\beta_t$ is aggregated additively and corresponds to scaling the volatility of the Brownian motion, which is captured by the square root transformation $\sqrt{2(\alpha_t+\beta_t)}$ (the factor of $2$ is just a convenient normalization without loss). 
    Instantaneous volatility here is the continuous-time analogue of increasing variance in a discrete-time setting. 
    In other words, players are always able to escalate instability, but they cannot decrease instability triggered by the opponent.  
    
    Observe that instability here is entirely endogenous: players can remain at the current status quo forever if they choose not to increase volatility ($\alpha_t=\beta_t=0$), but each player has the ability to unilaterally generate instability. 
    In this sense, a state $X_t$ at which no player has an incentive to generate instability corresponds to a stable status quo. 
    We focus on the stylized case in which {all} instability is endogenous to clearly identify its idiosyncratic effects. 

    Creating instability is costly. 
    We assume the cost of instability effort is convex and adopt a quadratic specification for simplicity. 
    The instantaneous (net) payoffs of $A$ and $B$ are respectively: 
    \begin{align*}
        u_a(X_t,\alpha_t) &:= X_t -  \frac{c_a}{2} \alpha_t^2,
        &
        u_b(X_t,\beta_t) &:= (1-X_t) -  \frac{c_b}{2}\beta_t^2,
    \end{align*}
    \noindent where $c_a,c_b \in \mathbb{R}_{++}$ are idiosyncratic cost parameters for each player. 
    
    Because creating instability is costly, this is not a zero-sum game. 
    At a given instant, instability requires {a pure destruction of surplus} which can only be warranted by the hope of obtaining a durably better situation in some appropriate sense. 

    Each player chooses its instability effort over an infinite horizon. 
    Players have discount factors $r_a$ and $r_b$ respectively; flow payoffs are normalized by the discount factors. 
    Expected utilities as a function of strategies and the status quo (the initial point of the process $X_0 = x$) are given by
    \begin{align*}
        U_a(\alpha,\beta\,\mid\,x) &:= \mathbb{E} \left[ \int_0^\infty  r_a e^{-r_a t} u_a(X_t,\alpha_t) d_t \right],
        &
        U_b(\alpha,\beta\,\mid\,x) &:= \mathbb{E} \left[ \int_0^\infty  r_b e^{-r_b t} u_b(X_t,\beta_t) d_t \right].
    \end{align*}
    
    We restrict attention to Markov-perfect equilibria \citep{TiroleMaskin2001JET} in continuous strategies. 
    We then denote strategies as $\alpha_t = a(X_t)$, $\beta_t=b(X_t)$, where $a$ and $b$ are continuous functions from $[0,1]$ to $\mathbb{R}_+$. 
    Formally, strategies belong to the class of $X_t$-adapted progressively measurable processes, which we denote by $\mathcal{A}$. 
    The restriction to Markov-perfect equilibria is common in the literature, due in part to well-known issues in defining off-path behavior in continuous time \citep[see][]{SimonStinchcombe1989Ecta}. 
    Continuity is partly a technical assumption, albeit a natural one in our setup. 
    It is also minimal in that it requires little regularity to ensure that the underlying objects are properly defined. 
    We formally define our equilibrium concept: 

    \begin{dfn}
    An equilibrium is a pair of continuous functions $(a,b)$ from $[0,1]$ to $\mathbb{R}_+$ 
    such that:
    \begin{enumerate}[label=(\roman*)]
        \item The process $\alpha_t^* = a(X_t)$ solves the control problem for player $A$ given $b$:
        \begin{align*}
            \alpha^* \in \argmax_{\alpha \in \mathcal{A}} \; & \mathbb{E} \left[ \int_0^\infty  r_a e^{-r_a t} \left( X_t -   \frac{c_a}{2}\alpha_t^2 \right) d_t \right]
            &
            &\, \text{ s.t. } dX_t = \sqrt{2(\alpha_t+b(X_t))}dB_t - dK_t, \,\,
            X_0 = x.
        \end{align*}
        \item The process $\beta_t^* = b(X_t)$ solves the control problem for player $B$ given $a$:
        \begin{align*}
           \beta^* \in \argmax_{\beta \in \mathcal{A}} \; & \mathbb{E} \left[ \int_0^\infty  r_b e^{-r_b t} \left( (1-X_t) -   \frac{c_b}{2}\beta_t^2 \right) d_t \right]
            &
            &\, \text{ s.t. } dX_t = \sqrt{2(a(X_t)+\beta_t)}dB_t - dK_t, \,\,
            X_0 = x.
        \end{align*}
    \end{enumerate}
    \end{dfn}

    In the next section, we study the control problem in detail for a fixed strategy of the opponent so as to characterize best responses in this game. 
    In doing so, we will verify that the previous definition of equilibrium is appropriate; in particular, optimal strategies are well-defined and continuous.  
    This also allows us to identify relevant properties of best responses, which will prove useful to provide a {direct construction of equilibria} in \hyref{sec:equilibrium}{Section}.

\section{Characterizing Best Responses}
\label{sec:controlproblem}

    In this section, we study the properties of players' best responses through its differential characterization.
    We consider the control problem of one player, holding fixed the strategy of the opponent. 
    Since the individual problems of the players are symmetric by definition when replacing $X_t$ by $1-X_t$ in the flow payoff, we will consider player $A$'s problem. 
    All results extend symmetrically to player $B$'s problem. 
    As we focus on player $A$'s problem, throughout this section we will omit the $a$ subscripts on parameters $r_a,c_a$ and instead write $r,c$ to alleviate notation. 

    To formally define the control problem that we study in this section, let $(\Omega,\mathcal{F},(\mathcal{F}_t),\mathbb{P})$ denote a complete filtered probability space equipped with a one-dimensional Brownian motion $B_t$. 
    Fix $b:[0,1] \rightarrow \mathbb{R}_+$ a continuous function.
    The control problem of the player is given by:
    \begin{align*}
           v_a(x)  = \sup_{ \alpha \in \mathcal{A} } \; & \mathbb{E} \left[  \int_0^\infty r e^{-r t} \biggl( X_t - \frac{c}{2}\alpha_t^2 \biggr) dt \right]
           &
           &\, \text{ s.t. } dX_t = \sqrt{2 \bigl(\alpha_t + b(X_t) \bigr) } dB_t - dK_t, \,\,
            X_0 = x,
    \end{align*}
    \noindent where $X_t,K_t$ solve the reflection problem i.e $X_t \in [0,1]$. 

    The following subsection introduces the approach used to solve the control problem: a differential characterization of the problem and the theory of viscosity solutions. 

    \subsection{Differential Characterization: Existence and Uniqueness} 
    \label{sec:controlproblem:diff-characterization}

    The value function of the control problem (and therefore the optimal control) is fully characterized as the solution to a second-order differential equation --- the Hamilton--Jacobi--Bellman (HJB) equation --- with reflective boundary conditions, which capture the fact that the status quo is reflected on a closed interval. 
    
    To state the main result of this subsection, denote the positive part of $y\in \mathbb R$ by the subscript $+$, $y_+:=\max\{y,0\}$, and let $n(x)$ denote the outer normal unit vector, where $n(0)=-1,n(1)=1$.
    \begin{thm}
        \label{thm:HJBBC-controlproblem}
        The value function $v_a$ is the unique viscosity solution to the following Hamilton--Jacobi--Bellman equation:
        \begin{align*}
               r v_a(x) - \sup_{a \in \mathbb{R}_+} \left\{  r x -  \frac{r c}{2}a^2 + \bigl( a + b(x) \bigr) v_a''(x)  \right\} = 0  & \quad  \text{ on } (0,1) \tag{HJB} \label{eq:HJB}
        \end{align*}
        with the reflective boundary condition:
        \begin{align*}
              n(x)v'_a(x) = 0 & \quad \text{ on } \{0,1\}. \tag{BC} \label{eq:BC}
        \end{align*}
        Furthermore, $v_a$ is continuous and, whenever $v_a''$ exists, the optimal control is given by
        \begin{align*}
            a(x) = \frac{1}{rc} v''_a(x)_+.
        \end{align*}
    \end{thm}
    \noindent We will refer to the combination of \hyperref[eq:HJB]{\color{black}(\ref*{eq:HJB})} and \hyperref[eq:BC]{\color{black}(\ref*{eq:BC})} as the \emph{reflected problem} \hyperref[eq:HJB]{\color{black}(RP)} given $b$. 
    
    The value function only solves \hyperref[eq:HJB]{\color{black}(RP)} in an appropriate weak sense we define below: it is a viscosity solution.\footnote{
        For general references on the theory of viscosity solutions of elliptic second-order differential equations and its applications to optimal control see  \citet{CrandallIshiiLions1992BullAMS}, \citet{FlemingSoner2006}, and \citet{Pham2009Book}.
    } 
    We rely on the use of viscosity solutions because a number of issues render our problem non-standard. 
    First, the process degenerates and becomes deterministic if there is no instability. 
    If for some $x \in [0,1]$ $b(x)=0$, then by setting $\alpha_t=0$ the player can make the process constant. 
    In particular, this implies the boundary conditions need not be satisfied as players could choose to `deactivate' the reflection by setting $a(x)=b(x)=0$ at $x \in \{0,1\}$. 
    Indeed if $b(1)=0$, it is immediate that player $A$ has no interest to generate instability when the status quo is 1, as $A$ enjoys the maximum possible payoff forever. 
    This effectively makes $1$ an absorbing point and the strong boundary condition fails to hold in the usual sense. 
    In general, whether or not the (strong) boundary conditions hold is tightly related to the activity of the other player. 
    
    Second, players' best responses generally do not satisfy standard regularity conditions (as Lipschitz continuity), which prevents us from appealing to well-known result for existence and uniqueness. 
    Finally, the value function can be non-differentiable; we will show below it can exhibit a kink in equilibrium. 
    The presence of a kink is more than a technical curiosity and will reflect essential properties of an equilibrium: a kink appears at a stable status quo that is supported but both players threatening to generate enough instability on either side to prevent any deviations. 
    We give more details below when characterizing the value function and equilibrium. 
    
    Before we introduce viscosity solutions, observe that the \hyperref[eq:HJB]{\color{black}(HJB)} equation can be rewritten as:
    \begin{align*}
        r v_a(x) - r x - b(x) v_a''(x) - \frac{1}{2rc}\bigl[v_a''(x)_+\bigr]^2 = 0.
    \end{align*}
    For convenience, we define the following notation for the differential operators: 
    \begin{align*}
        F_a(x,v,M) &:= r v - r x - b(x) M - \frac{1}{2rc}\bigl[M_+\bigr]^2 && \quad \text{for } (x,v,M) \in [0,1] \times [0,1] \times \mathbb{R}
        \\
        B(x,p) &:= n(x) p && \quad \text{for } (x,p) \in \{0,1\} \times \mathbb{R}
    \end{align*}
    that is, \hyperref[eq:HJB]{\color{black}(RP)} is given by $F_a(x,v_a(x),v''_a(x))=0$ on $(0,1)$ and $B(x,v'_a(x))=0$ on $\{0,1\}$. 
    We now state the definition of a viscosity solution of \hyperref[eq:HJB]{\color{black}(RP)}:
    \begin{dfn}
        A function $w$ on $[0,1]$ is a viscosity \textbf{subsolution} of \hyperref[eq:HJB]{\color{black}(RP)} if its upper-semicontinuous envelope $w^*$ is satisfies 
        \begin{align*}
            F_a(x_0,w^*(x_0),\varphi''(x_0))  &\leq 0 \text{ if } x_0 \in (0,1)
            \\ 
            \, \text{ and } \, 
            \min \bigl\{  F_a(x_0,w^*(x_0),\varphi''(x_0)), B(x_0,\varphi'(x_0)) \bigr\} & \leq 0 \text{ if } x_0 \in \{0,1\}
        \end{align*}
        for all $\varphi \in \mathcal{C}^2([0,1])$ such that $x_0$ is a local \textbf{maximum}  of $w^*-\varphi$.
        
        A function $w$ on $[0,1]$ is a viscosity \textbf{supersolution} of \hyperref[eq:HJB]{\color{black}(RP)} if its lower-semicontinuous envelope $w_*$ is satisfies 
        \begin{align*}
            F_a(x_0,w_*(x_0),\varphi''(x_0)) & \geq 0 \text{ if } x_0 \in (0,1)
            \\
            \, \text{ and } \,
            \max \bigl\{  F_a(x_0,w_*(x_0),\varphi''(x_0)), B(x_0,\varphi'(x_0)) \bigr\} & \geq 0 \text{ if } x_0 \in \{0,1\}
        \end{align*}
        for all $\varphi \in \mathcal{C}^2([0,1])$ such that $x_0$ is a local \textbf{minimum}  of $w_*-\varphi$.

        A function $w$ is a \textbf{viscosity solution} if it is both a viscosity sub- and supersolution. 
    \end{dfn}
    
    \noindent Viscosity solutions provide a powerful notion of generalized differentiability which is well adapted to studying HJB-type equations. 
    One canonical intuition to visualize the viscosity approach is to think about fitting smooth test functions --- $\varphi$ in the definition --- equal to the function at a given point but everywhere else above (for a subsolution) or below (for a supersolution) and requiring the differential equation to hold with the appropriate inequality for any such test function. 

    The proof of \hyref{thm:HJBBC-controlproblem}{Theorem} is a combination of two propositions (proved in the Appendix).
    First, we prove that the value function is a viscosity solution to the stated equation.
    \begin{prop}[Optimality]
        \label{prop:optimality-controlproblem}
        The value function $v_a$ is a viscosity solution to \hyperref[eq:HJB]{\color{black}(RP)}.
    \end{prop}

    \noindent \hyref{prop:optimality-controlproblem}{Proposition} follows from standard dynamic programming arguments and applying Ito's lemma to appropriate test functions, although our setup imposes minimal assumptions. 

    We then turn to proving we have a unique viscosity solution, therefore corresponding to the value function itself. 
    To do so, we first establish a comparison principle result that will also be of practical interest in characterizing equilibrium properties.

    \begin{lemma}[Comparison Principle]
        \label{lemma:comparisonprinciple-controlproblem}
        If $\overline{w}$ is a viscosity supersolution and $\underline{w}$ is a viscosity subsolution to \hyperref[eq:HJB]{\color{black}(RP)}, then $\overline{w} \geq \underline{w}$ in $[0,1]$.
    \end{lemma}

    \noindent The comparison principle allows us to find bounds for our solution by constructing sub- and supersolutions.
    Moreover, since existence can be established using general arguments, the comparison principle is instrumental in proving uniqueness.\footnote{
        To prove existence, it is sufficient to exhibit a subsolution (take $\underline v(x):=0$) and a supersolution (take $\overline v(x):=1$) such that the latter is everywhere above the former. 
        We can then construct a solution by taking the pointwise supremum of subsolutions that are everywhere below that supersolution. 
        This is known as Perron's methods in the viscosity solution literature. 
        The comparison principle then immediately implies uniqueness of a viscosity solution.
    }

    \begin{prop}[Existence and Uniqueness]
        \label{prop:existenceuniqueness-controlproblem}
        There exists a unique viscosity solution to \hyperref[eq:HJB]{\color{black}(RP)}. Furthermore, it is continuous.
    \end{prop}

    \noindent The proof of \hyref{lemma:comparisonprinciple-controlproblem}{Lemma} and \hyref{prop:existenceuniqueness-controlproblem}{Proposition} relies on adapting existing techniques from the literature \citep[see][]{CrandallIshiiLions1992BullAMS} with arguments that are idiosyncratic to the problem at hand.

    \subsection{Properties of Best Responses} 
    \label{sec:controlproblem:properties-br}

    We now characterize player $A$'s optimal control for an arbitrary strategy by player $B$.
    Recall that player $A$'s control is characterized by
    \begin{equation*}
    a(x) = \frac{1}{rc} v''_a(x)_+,
    \end{equation*}

    Even if player $A$'s optimal control does depend on $b$ (the dependence of $b$ is encoded within the value function $v_a$), the following theorem shows that the best response and value function always exhibit a simple structure characterized by a threshold mechanism. 

    \begin{thm}{(Best Response Characterization)}
    \label{thm:best-response-characterization}
    Let $b$ be a continuous function and $v_a$ the solution to problem \hyperref[eq:HJB]{\color{black}(RP)} given $b$. 
    The optimal control $a^*$ exists, and the solution to the control problem is fully characterized by two thresholds $\underline{x}_a,\overline{x}_a \in (0,1]$, $\underline{x}_a \leq \overline{x}_a$, such that:
    \begin{enumerate}[label=(\roman*)]
        \item on $[0,\underline x_a)$ (\emph{the beneficial instability region}), $v_a$ is strictly convex, and strictly above the identity;
        \item on $[\underline{x}_a,\overline{x}_a]$ (\emph{the neutral region}), $v_a(x)=x$;
        \item on $(\overline x_a,1]$ (\emph{the detrimental instability region}), $v_a$ is strictly concave, and strictly below the identity;
        \item $a^*$ is continuous, strictly positive on $[0,\underline x_a)$ and zero elsewhere.
    \end{enumerate}
    Furthermore, $v_a$ is increasing and twice continuously differentiable everywhere, except possibly at $\overline{x}_a$.
    If $v_a$ is not differentiable at $\overline{x}_a<1$, then  $\lim_{x \rightarrow \overline{x}_a^-} v_a'(x) \geq \lim_{x \rightarrow \overline{x}_a^+} v'_a(x)$ (only concave kinks are permissible).
    \end{thm}
    \noindent
    Let us discuss the intuition underlying \hyref{thm:best-response-characterization}{Theorem} and its implications.

    First, observe that the threshold structure is a general feature of best responses, regardless of player $B$'s strategy. 
    The optimal strategy for $A$ always consists of generating strictly positive instability when the status quo is unfavorable enough, and doing so in a vanishing manner as the player reaches a `satisficing' threshold $\underline{x}_a$. 
    The fact that $v_a$ is strictly above the identity in the beneficial instability region captures the idea that $A$ is strictly better off there than if they were able to stay at that status quo forever. 
    Further, the fact that $v_a$ is convex in this region captures the (positive) option value from instability. 
    This option value decreases as player $A$'s share nears $\underline{x}_a$ and the player becomes more prudent as they have more to lose. 

    The lower threshold $\underline{x}_a$ synthesizes player $A$'s ability to use instability to their advantage and is determined both by $b$, the discounting factor, and the cost parameter. 
    Essentially, expected gains from instability come from durably experiencing more favorable states.
    Beyond $\underline{x}_a$, it would be too costly or not beneficial enough to try to generate instability in their favor.
    This can be because it would require too long a span of instability --- entailing too high a cost --- to exploit the option value offered by the lower bound and secure durable improvements, or because player $B$ would generate enough instability at more favorable situations for player $A$ so as to prevent them from durably improving their situation there.

    The beneficial instability region $[0,\underline x_a)$ is always non-empty: 
    there is always a benefit to generating some instability when the status quo is too disadvantageous. 
    Reflection binds at the lower bound, where the player has the worst possible payoff; for any interior status quo, instability is locally equally likely to make the player worse off or better off. 
    Yet, the fact that there is a worst state generates option value and the strict incentive to increase volatility at the bound spills over and makes it profitable to generate instability {in a nearby region}. 
    Such a threshold not only always exists for arbitrary $b$, but it is also always strictly above zero, which demonstrates that instability always enables players to fight off against situations that are too unfavorable. 

    The upper threshold $\overline{x}_a$ only matters for determining the payoff structure at states in which player $A$ does not contribute to instability. 
    It delineates an intermediate {neutral region} $[\underline x_a,\overline x_a]$ where, even though player $A$ chooses not create instability, whatever instability might be generated by the opponent is not harmful (expected payoff are equal to status quo payoffs).
    For states that are strictly preferred to $\overline x_a$ by player $A$ --- where they have a lot to lose --- whatever instability is generated by the opponent is actively harmful to player $A$.
    This is captured by the fact that $v_a(x)<x$ in this region: player $A$ would prefer staying at the status quo, and $v_a$ is concave as the option value of instability is negative.
    Although the neutral region is always non-empty (but possibly consisting of a single point), the detrimental instability region can be empty.
    Additionally, in general, it need not be the case that $b(x)=0$ in the neutral region.
    
    A corollary of the previous results is that $\overline{x}_a<1$ if and only if $b(1)>0$. 
    This highlights how instability at the extreme states significantly influences the player's behavior in the interior of the domain. 
    If $b(1)=0$, no amount of instability that player $B$ otherwise generates inside the domain can be harmful to $A$ (the detrimental instability region is empty). 
    As in this case $b$ entails no instability at player $A$'s preferred state ($x=1$), and as player $A$ would never optimally generate instability at this most favorable status quo, $A$ is then able to make this boundary fully absorbing. 
    What happens at the extremes has drastic consequences everywhere else: whatever instability $B$ otherwise generates is non-harmful and instead benefits player $A$. 
    On the other hand, if the boundary at $1$ is actively reflecting ($b(1) > 0$), player $A$ cannot unilaterally stop the process at $1$, and the option value of increasing instability when close to 1 is negative.

    Lastly, we provide a monotonicity result that suggests what will be the structure of equilibria: if one player creates more instability when more disadvantaged, but becomes more conservative as the status quo is more favorable to them, the other player will have incentives to do the same.
    \begin{prop}\label{prop:decreasing-control}
        If $b$ is non-decreasing on $[0,\underline x_a]$, then the optimal control to the problem \hyperref[eq:HJB]{\color{black}(RP)} is non-increasing.
    \end{prop}

    \subsection{The Inactive Benchmark}\label{sec:controlproblem:inactive-benchmark}

    Consider the case in which a player's opponent is fully passive and never generates any instability. 
    We take player $A$'s viewpoint, with $b(x)\equiv 0$ for all $x$, so that player $A$'s actions are the only source of instability to the status quo. 
    The analysis of this individual decision-making problem not only serves to ground intuition, but most importantly, key properties of equilibria will be determined by what would happen in the inactive benchmark.

    The HJB equation has a clear interpretation: it relates the instantaneous cost of control at the optimum to the marginal benefit relative to the status quo, which can be seen as the option value of instability. 
    This also highlights why the second-order derivative in this context captures the option value.
    Indeed, rewrite the HJB as:
    \begin{align*}
        \underbrace{r (v_a(x) - x)}_{\text{improvement on the status quo}} = \underbrace{ \frac{1}{2rc} [v_a''(x)_+]^2 }_{\text{option value}} = \underbrace{\frac{rc}{2} a^*(x)^2}_{\text{instantaneous cost of control}}
    \end{align*}
    The next proposition strengthens \hyref{thm:best-response-characterization}{Theorem} when restricting to the special case $b\equiv 0$. 
    \begin{prop}[Properties of the Inactive Benchmark]
    \label{prop:properties-individual-control}
    Let $v_a^0$ be the value function in \hyperref[eq:HJB]{\color{black}(RP)} given $b \equiv 0$, and $a^{*,0}$ be the corresponding optimal control.
    Then, there is $\underline{x}_a^0 \in (0,1]$ such that
     \begin{enumerate}[label=(\roman*)]
        \item  on $[0,\underline{x}_a^0)$, $v_a^0$ is {strictly} convex, $v_a^0(x)>x$, and $a^{*,0}$ is strictly positive and strictly decreasing; 
        \item on $[\underline{x}_a^0,1]$, $v_a^0(x)=x$ and $a^{*,0}(x)=0$.
    \end{enumerate}
    Moreover,
    $v_a^0$ and $a^{*,0}$ are twice-continuously differentiable except possibly at $\underline x_a^0=1$, with $v_a^{0\prime}\leq 1$.
    \end{prop}

\begin{figure}[t]
        \centering
        \makebox[\textwidth][c]{
        \begin{minipage}{1.0\linewidth}
            \begin{subfigure}{.5\linewidth}
                \includegraphics[width=1\linewidth]{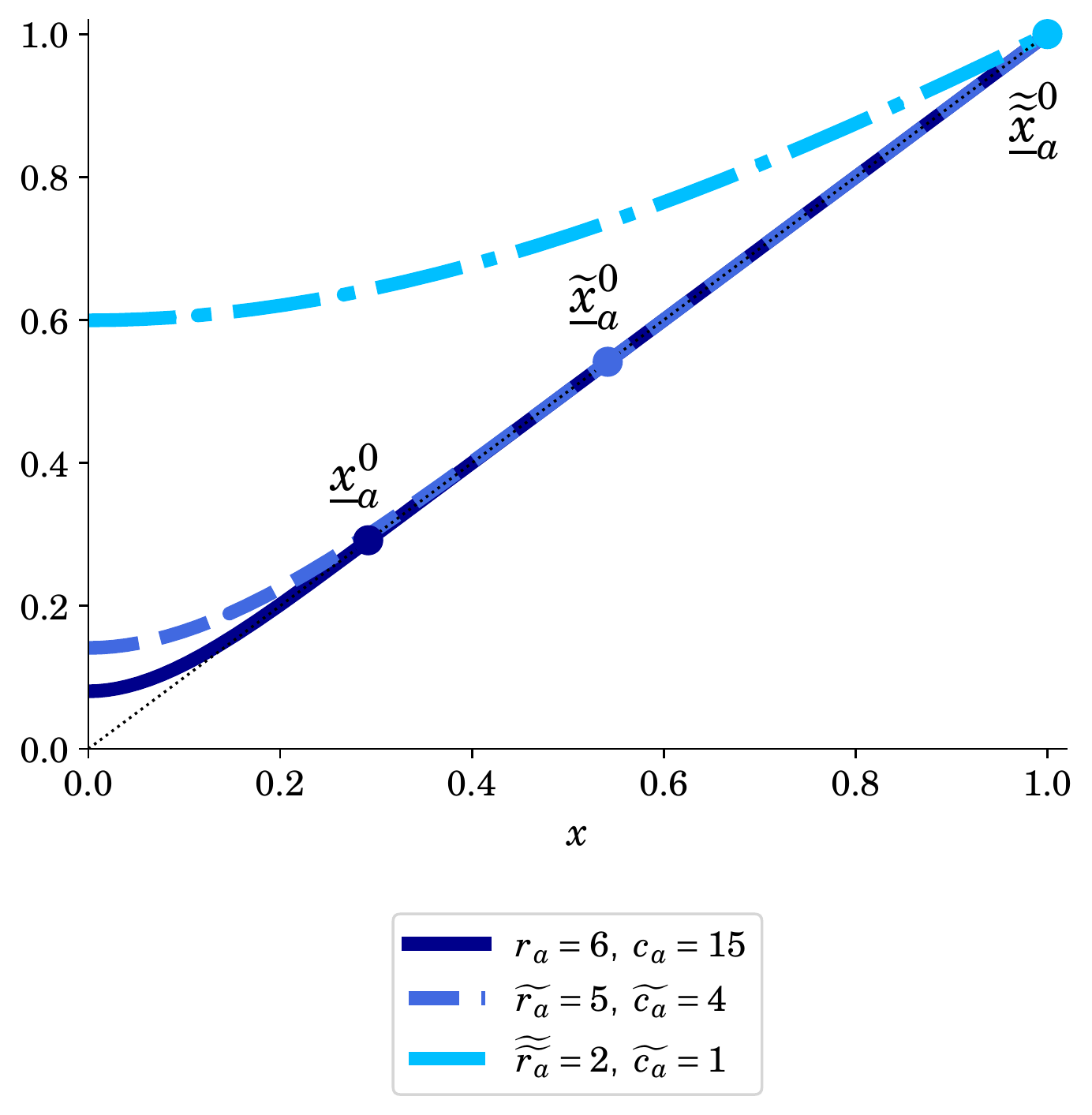}
                \caption{Value Function}
                \label{fig:idmplotAcs:value}
            \end{subfigure}
            \begin{subfigure}{.5\linewidth}
                \includegraphics[width=1\linewidth]{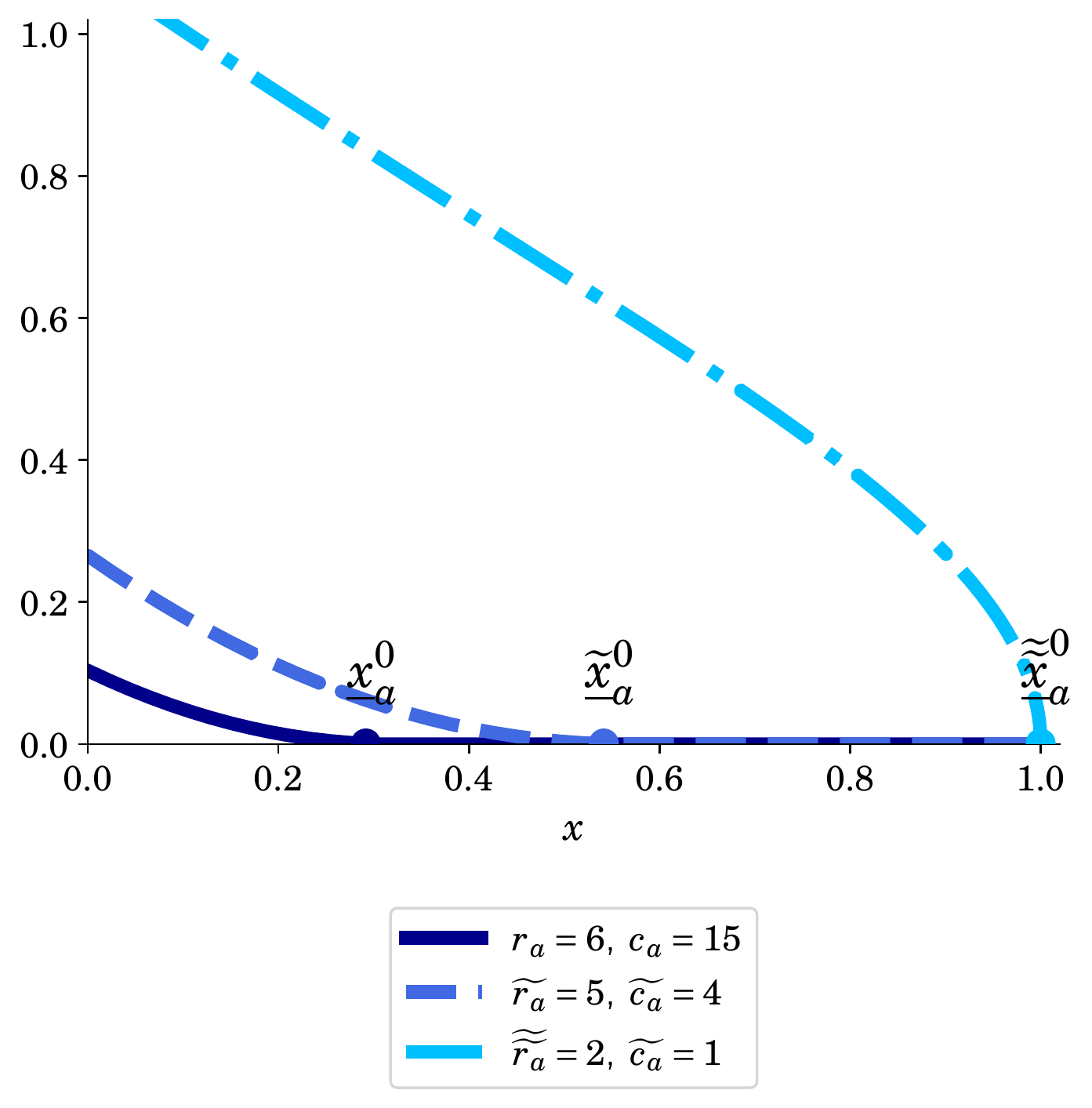}
                \caption{Optimal Strategy}
                \label{fig:idmplotAcs:strat}
            \end{subfigure}
        \end{minipage}
        }
        \caption{Comparative Statics for Player $A$ in the Inactive Benchmark}~\\
        \label{fig:idmplotAcs}
    \end{figure}

    \hyref{fig:idmplotAcs}{Figure} illustrates \hyref{prop:properties-individual-control}{Proposition} with a numerical approximation of the value function and the optimal control of player $A$ in the inactive benchmark for different parameter values. 
    It exhibits the typical best-response structure: the value function is convex and above the identity when $x$ is low enough; it meets the identity at $\underline x_a^0$, and remains at the status quo for greater values of $x$.
    
    Since $b\equiv 0$, the detrimental instability region is empty:
    if player $A$ fully controls instability, then they will never choose harmful levels of instability as they can always guarantee the status quo --- therefore $v_a^0(x) \geq x$ everywhere.
    Moreover, player $A$'s inaction region $[\underline{x}_a^0,1]$ determines the states at which, for the given cost and discounting parameters, player $A$ has {no possible intrinsic benefit from instability}. 
    The active region $[0,\underline{x}_a^0)$ symmetrically delineates the situations where $r$ and $c$ are such that player $A$ {can} strictly profit from instability. 
    By the same logic, the fact that $a^{*,0}$ is {strictly decreasing} over the active region $[0,\underline{x}_a^0)$ captures the idea that the return to instability is decreasing as the status quo moves farther away from zero: at more favorable states the improvement on the status quo shrinks and so does the value to generating instability. 
    \hyref{fig:idmplotAcs:strat}{Figure} also depicts the corresponding optimal control to \hyref{fig:idmplotAcs:value}{Figure}, illustrating this decreasing behavior.

    We further provide comprehensive comparative statics on $r,c$ which will later prove useful to obtain comparative statics of equilibria.
    Rewrite the HJB equation as $2 r^2 c (v_a(x)-x) - [v_a''(x)_+]^2 = 0$.
    Assume $v_a$ solves this equation, along with the boundary condition (in the viscosity sense) for $r,c$, and denote $\underline{x}_a$ the corresponding inaction threshold.
    Let $\tilde{r},\tilde{c}$ such that $\tilde{r}^2\tilde{c} \geq r^2 c$. Directly $2 \tilde{r}^2 \tilde{c} (v_a(x)-x) - [v_a''(x)_+]^2 \geq 0$.
    Since the boundary conditions are still verified in the viscosity sense, we can conclude that $v_a$ is a supersolution in the problem for $\tilde{r},\,\tilde{c}$. 
    Comparative statics then follow from the comparison principle (\hyref{lemma:comparisonprinciple-controlproblem}{Lemma}); they are summarized in the next proposition:
    \begin{prop}
    \label{prop:idm-comparative-statics}
       Consider two pairs of cost and discounting parameters $r,c$ and $\tilde{r},\tilde{c}$. 
       Denote $v_a^0$, ${x}_a^0, a^{*,0}$ the value function, optimal threshold, and control corresponding to the problem for $r,c$. 
       Similarly define $\tilde{v}_a^0$, $\tilde{x}_a^0$, $\tilde{a}^{*,0}$, ${a}^{*,0}$ under $\tilde{r},\tilde{c}$. 
       If $\tilde{r}^2 \tilde{c} > r^2 c$, then,
       \begin{enumerate}[label=(\roman*)]
           \item $\tilde{x}_a^0 \leq {x}_a^0$, with strict inequality if $\tilde x_a^0<1$;
           \item $\tilde{a}^{*,0} \leq {a}^{*,0}$ on $[0,1]$, with strict inequality on $[0,x_a^0)$; and
           \item $\tilde{v}_a^0 \leq v_a^0$ on $[0,1]$, with strict inequality on $[0,x_a^0)$.
       \end{enumerate}
    \end{prop}

    \noindent The interpretation of \hyref{prop:idm-comparative-statics}{Proposition} is quite natural. 
    For a higher cost/impatience, instability is less profitable overall. 
    Since the option value of generating instability is fully due to the player's forward-looking behavior, higher impatience reduces the option value provided by the lower bound on the state $x$. 
    Higher $c$ raises the marginal cost of instability. 
    As a consequence, higher cost or impatience cause the region where it is beneficial to generate instability to shrink ($\tilde{x}_a^0 \leq {x}_a^0$). 
    The instability generated at any state $x$ is milder ($\tilde{a}^{*,0} \leq {a}^{*,0}$), resulting in lower payoffs ($\tilde{v}_a^0 \leq v_a^0$).

    As a player becomes more patient and faces lower costs to instability, the player may find it worthwhile to generate strictly positive instability everywhere but at 1, i.e. $\underline x_a^0=1$. 
    The player's threshold is also associated with the shape of the player's optimal instability as illustrated in \hyref{fig:idmplotAcs:strat}{Figure}.
    When player $A$ stops instigating instability at $\underline x_a^0<1$, then the fact that the flow payoffs are bounded above by 1 ($x$ denotes a share) never comes into play: the upper bound is inactive.
    In this case, the player's optimal control exhibits a convex shape, and the instability generated by player $A$ vanishes smoothly, with ${a^{*,0}}'(x)\to 0$ as $x\to \underline x_a^0$, as observed from the darker solid and dashed lines in \hyref{fig:idmplotAcs:strat}{Figure}.
    In contrast, we observe that if the discount rate or the cost to instability are low enough, the player adjusts volatility to exactly attain its first best and avoid the upper bound becoming actively reflecting.
    Then, the player only stops generating instability exactly at $\underline x_a^0=1$, and we obtain the convex-concave shape for $a^{*,0}$ that we observe in the dashed-dotted line in \hyref{fig:idmplotAcs:strat}{Figure}, associated to instability vanishing abruptly at $\underline x_a^0$.
    We show that this distinction between the cases depicted in \hyref{fig:idmplotAcs:strat}{Figure} is in fact a generic property:
    
    \begin{prop}[Implications of Active Upper Bound]
    \label{prop:properties-upper-reflection}
        Let $v_a^0$ be the value function in \hyperref[eq:HJB]{\color{black}(RP)} given $b \equiv 0$, and $a^{*,0}$ be the corresponding optimal control.
        Then,
        \begin{enumerate}[label=(\roman*)]
            \item $a^{*,0}$ is convex if and only if $v_{a,-}^{0\prime}(\underline x_a^0)=1$;
            \item there is $\hat x_a\in [0,\underline x_a^0)$ such that $a^{*,0}$ is convex on $[0,\hat x_a]$ and concave on $[\hat x_a,\underline x_a^0]$ if and only if $v_{a,-}^{0\prime}(\underline x_a^0)<1$.
        \end{enumerate}
        Furthermore, $v_{a,-}^{0\prime}(\underline x_a^0)=1$ if $\underline x_a^0<1$, and $\lim_{x\uparrow \underline x_a^0}a^{*,0\prime}(x)= -\infty$ if $v_{a,-}^{0\prime}(\underline x_a^0)<1$.
    \end{prop}
    
    All the results in this section carry symmetrically to player $B$'s problem, with the change of variable $y=1-x$ (inverting the interval by relabeling $1$ as $0$ and vice versa). 
    This follows from symmetry of the problem, since player $B$'s best response problem is exactly player $A$'s problem after the change of variable --- up to possibly heterogeneous parameters $r_b,c_b$. 
    Equivalently, this amounts to expressing everything in terms of the payoffs to player $B$ instead of player $A$.

\section{Characterizing Equilibria}
\label{sec:equilibrium}
We now turn our attention to characterizing (Markov-perfect) equilibria of the game.
These will be pairs of strategies $(a,b)$ such that each is a best response to the other, and so any equilibrium will necessarily have to comply with the properties discussed in the previous section.
In this subsection, we first establish three necessary properties inherent to any equilibrium: (i) at most one player is creating instability, (ii) the more favorable the status quo for a player, the lower the instability that player generates, and (iii) each player generates lower instability than they otherwise would were their opponent passive.

This characterization allows us to delineate two possible cases for equilibrium, depending on whether the instability regions in each player's inactive benchmark overlap.
If they do not overlap, this gives rise to a unique \emph{accommodating equilibrium}, where players take an accommodating attitude toward the pursuit of a more favorable status quo by their opponents, generating many stable states.
If they do overlap, this leads to the existence of multiple \emph{deterrence equilibria}, each characterized by a unique stable status quo that is sustained by a deterrence mechanism.

\subsection{Necessary Properties of Equilibria}
    
We first turn to a crucial property of \emph{equilibria}: the fact that equilibrium instability strategies decouple --- at most one player generates instability at any given status quo.
Note that this feature is not immediately implied by our characterization of individual best-responses in \hyref{sec:controlproblem:properties-br}{Section}: there are strategies $b$ for which player $A$'s best response involves generating instability at states $x$ for which $b(x)>0$.
Nevertheless, any equilibrium of the game is uniquely characterized by two thresholds that delineate three regions: a stable region, a region where only player $A$ generates instability, and a region where only player $B$ generates instability.
The next proposition summarizes those properties and characterizes the structure of equilibria.

\begin{prop}
\label{prop:equilibrium-properties}
In any equilibrium, there exist $\underline x, \overline x \in (0,1)$, $\underline x\leq \overline x$ such that
\begin{enumerate}[label=(\roman*)]
    \item $\forall x \in [0,\underline x)$, $a(x)>0=b(x)$;
    \item $\forall x \in (\overline x,1]$, $a(x)=0<b(x)$; and
    \item $\forall x \in [\underline x,\overline x]$, $a(x)=0=b(x)$.
\end{enumerate}
Furthermore, $a$ (resp. $b$) is strictly decreasing on $[0,\underline x)$ (resp. increasing on $(\overline x,1]$), and the equilibrium is uniquely pinned down by $\underline x$, $\overline x$.
\end{prop}
\noindent The first thing to note is that \hyref{prop:equilibrium-properties}{Proposition} distinguishes between states that trigger instability and those at which stability is attained.
The former are those that are deemed excessively unfavorable by either player $A$ --- $x \in [0,\underline x)$ --- or player $B$ --- $x \in (\overline x,1]$.

The argument for why this `decoupling' structure of equilibria emerges is simple: owing to the fact they have diametrically opposed interests (constant-sum gross flow payoffs), it is not possible that both players expect to strictly improve on the same status quo.
At most one of the players sees an advantage to generating instability at any given status quo.
We know that at extremes states, $x =0$ and $x=1$, the disadvantaged player will actively push back by creating instability.
After all, they have nothing to lose and, while costly, instability can only improve their situation.
Then, due to the fact that \emph{for any} strategy of their opponent the set of states at which they find it profitable to generate instability is convex and includes their most unfavorable state (as shown in \hyref{thm:best-response-characterization}{Theorem}), we obtain the existence of these three regions.

Second, \hyref{prop:equilibrium-properties}{Proposition} tells us as players benefit from a larger share of the available benefits, they become more conservative in how much volatility they create.
Recall that the benefit to instability derives solely from the option value provided by the finiteness of resources being shared, as the is no immediate gain to instability when $x \in (0,1)$.
However, as there is a natural lower bound on how unsatisfactory the outcome can be, patient players may want to take a calculated risk to reap the benefits of this option value.
The proof follows from the fact that equilibrium strategies exhibit this decoupled structure, combined with the fact that if the opponent is unresponsive to instability, the optimal control is monotone in the state just as in the inactive benchmark discussed in \hyref{prop:properties-individual-control}{Proposition}.

The next property of equilibria relates the players' equilibrium strategies with their optimal instability strategy in the inactive benchmark case.
While it is tempting to think that in general the player always attains the highest expected payoff when their opponent is passive ($b\equiv 0$), this is not the case.
Player $B$ could potentially take $A$'s stead in generating optimal instability and saving $A$ the cost of doing so.
However, in equilibrium, it is indeed true that player $A$ cannot be better-off than if facing a passive opponent:
\begin{prop}
\label{prop:individualcontrol-supersolution}
    In any equilibrium, $v_a\leq v_a^0$ and $v_b\leq v_b^0$.
\end{prop}
\noindent The result derives from two observations.
First, that $v_a$ is a subsolution to \hyperref[eq:HJB]{\color{black}(RP)} in the inactive benchmark case, for which $v_a^0$ is a solution.
Second, from the fact that, from the comparison principle (\hyref{lemma:comparisonprinciple-controlproblem}{Lemma}), we know that any subsolution is weakly smaller than a supersolution --- and thus, than a solution. 
This implies we can compare equilibrium thresholds to inactive benchmark thresholds, as well as equilibrium instability to inactive benchmark strategies ($a^{*,0},b^{*,0}$).
Specifically, a player will never generate more instability at any point than they would in the inactive benchmark. 
This observation is formalized in the following corollary:

\begin{corol}
     In any equilibrium with thresholds $\underline x,\overline x$ and equilibrium strategies $a^*,b^*$, (i) $\underline x\leq \underline x_a^0$ and $\underline x_b^0\leq \overline x$; and (ii) $a^* \leq a^{*,0}$ and $b^* \leq b^{*,0}$.
     Furthermore, $a^* < a^{*,0}$ and $b^* < b^{*,0}(x)$ on $[0,x_a^0)$ and $(x_b^0,1]$, respectively, if and only if, $\underline x < \underline x_a^0$ and $\underline x_b^0< \overline x$.
\end{corol}

What a player could and would do if their opponent were to play passively determines the structure of equilibrium.
\emph{In equilibrium} the optimal strategy $b$ of the opponent will never be beneficial to player $A$ because they have diametrically opposed interests: it is not possible that both players simultaneously benefit from instability in equilibrium.
In other words, the intuition that if instability is not beneficial at a given point when $b\equiv 0$, it is still not beneficial when $b \neq 0$ is \emph{true in equilibrium}, but not in general.
Additionally, the optimal strategy of the inactive benchmark and the inaction threshold $\underline x_a^0$ in particular can be interpreted as the players' ability to threaten their opponent.
This underpins the argument that the inactive case discussed in \hyref{sec:controlproblem:inactive-benchmark}{Section} is indeed the right benchmark.

While \hyref{prop:equilibrium-properties}{Propositions} and \ref{prop:individualcontrol-supersolution} deliver necessary properties of any equilibrium, they are silent about the existence of equilibria.
The remainder of this section is devoted not only to showing their existence, but also to further specializing the characterization of equilibria by delineating the two possible kinds of equilibrium, which depend on parameter values --- and in particular on the relative positions of $\underline x_a^0$ and $\underline x_b^0$.

\subsection{Characterization of Equilibria: Deterrence and Accomodation} 

The main result of this subsection fully characterizes equilibria of the game; it delineates two possible cases depending on the relative position of the thresholds in the inactive benchmark. 
When profitable instability regions in the inactive benchmark do not overlap, there is a unique accommodating equilibrium where both players follow their inactive benchmark strategies and there are many stable states ($\underline x < \overline x$). 
If profitable instability regions do overlap in the inactive benchmark, there are multiple deterrence equilibria each characterized by a single stable state $\underline x = \overline x$.
Then, equilibrium strategies can be obtained by solving an ``as if'' inactive benchmark on each player's respective restricted interval --- $[0,\overline x]$ for $A$, $[\overline x,1]$ for $B$ --- as if reflection occurred at $\overline x$. 
We first state the theorem, and then elaborate on its intuition and implications.

\begin{thm}
\label{thm:eqm-characterization}
    An equilibrium $(a^*,b^*)$ exists.
    
    \emph{(Accommodating equilibrium)} If $\underline x_a^0\leq \underline x_b^0$, there is a unique equilibrium given by $(a^*,b^*)=(a^{*,0},b^{*,0})$.
    Moreover, at any equilibrium such that $\underline x<\overline x$ it must be the case that $(\underline x,\overline x)=(\underline x_a^0,\underline x_b^0)$ .

    \emph{(Deterrence equilibrium)} If $\underline x_a^0>\underline x_b^0$, a pair of strategies $(a^*,b^*)$ is an equilibrium if and only if 
    $a^*(x)=\mathbf 1_{(x<\overline x)}\frac{1}{r_a c_a}v_a''(x)$ and 
    $b^*(x)=\mathbf 1_{(x>\overline x)}\frac{1}{r_b c_b}v_b''(x)$,
    where 
    $v_a$ and $v_b$ are the unique viscosity solutions to the respective inactive benchmark problems on $[0,\overline x]$ and $[\overline x,1]$, and 
    $\overline x \in [\underline x_b^0,\underline x_a^0]\setminus \{0,1\}$.
\end{thm}
\noindent Whenever there is a status quo such that neither of players wants to increase instability even if their opponent were passive ($\underline x_a^0\leq \underline x_b^0$), then equilibrium behavior is everywhere as if their opponent were indeed passive. Each player is accommodating towards their opponent's aspirations to obtain a better outcome for themselves by creating some instability; players never `push back' against one another.

It is worth emphasizing that an accommodating equilibrium, when it exists, must be the unique equilibrium.
In other words, when the inactive benchmark is such that there is no status quo where both players would be willing to generate instability if the other were inactive, then this is the only equilibrium outcome.

\begin{figure}
    \centering
    \makebox[\textwidth][c]{
    \begin{minipage}{1.0\linewidth}
        \begin{subfigure}{.5\linewidth}
            \includegraphics[width=1\linewidth]{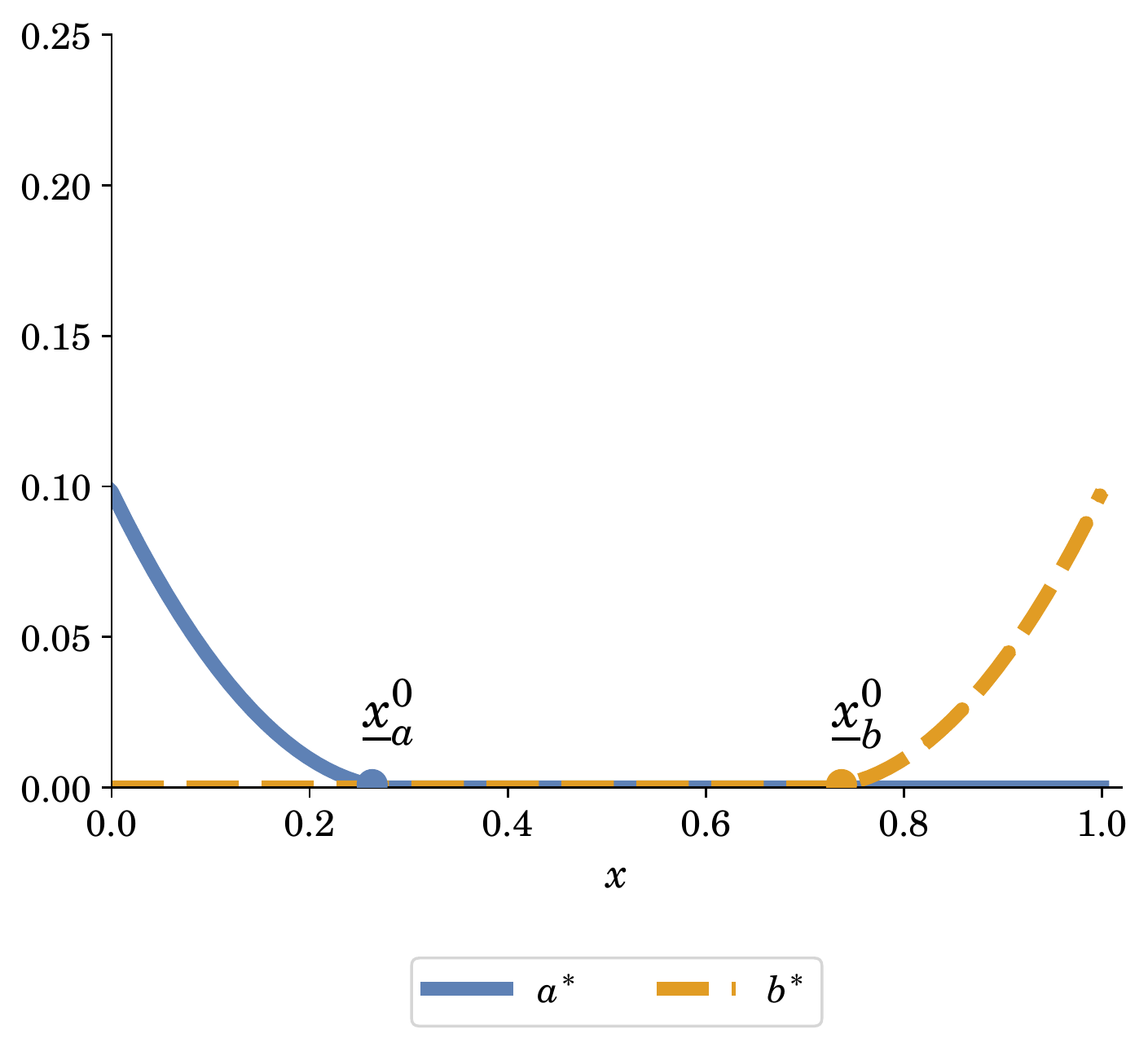}
            \caption{~}
            \label{fig:strat-AEqm:1}
        \end{subfigure}
        \begin{subfigure}{.5\linewidth}
            \includegraphics[width=1\linewidth]{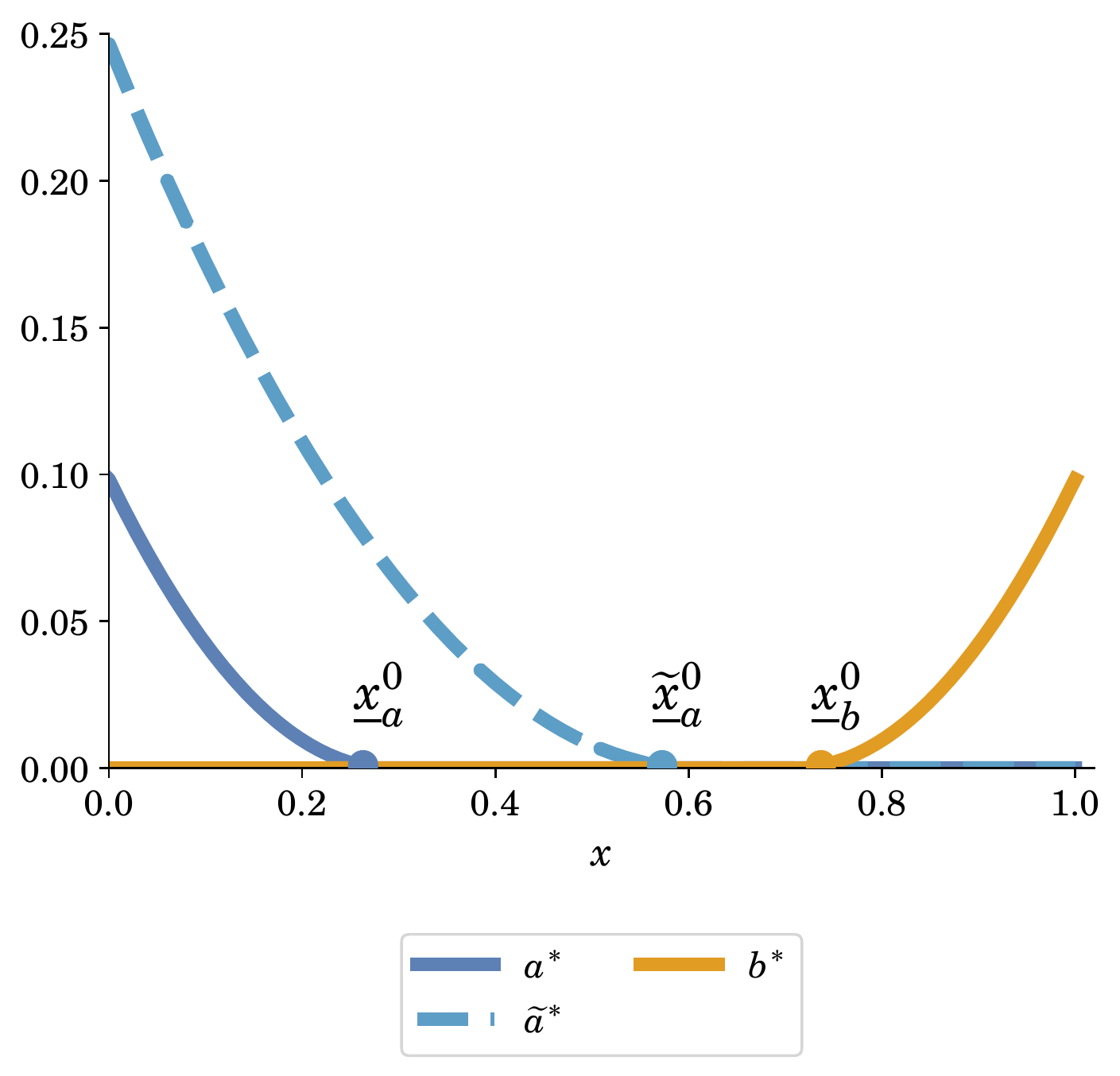}
            \caption{~}
            \label{fig:strat-AEqm:2}
        \end{subfigure}
    \end{minipage}
    }
    \caption{Equilibrium Strategies in an Accommodating Equilibrium}~\\
    \begin{minipage}{.95\linewidth}
        \footnotesize Note: Parameter values are $r_a=r_b=7$ and $c_a=c_b=15$ for solid lines, and $\widetilde{r_a}=4$, $\widetilde{c_a}=5$ for the dashed line.
    \end{minipage}
    \label{fig:strat-AEqm}
\end{figure}

Another noteworthy feature of accommodating equilibria is that they exhibit an interval $[\underline x_a^0, \underline x_b^0]$ of stable states where the status quo prevails.
This interval can be very large, as in \hyref{fig:strat-AEqm:1}{Figure} where approximately every state between $1/4$ and $3/4$ is stable.
There, the inability to profitably generate instability means that both players are willing to accept a large range of states.
As a consequence, states that can be potentially much more strongly preferred by one player than another can be sustained in the long run.

Given that, from \hyref{prop:idm-comparative-statics}{Proposition}, $\underline x_a^0$ (resp. $\underline x_b^0$) is decreasing (resp. increasing) with respect to $r_a$ and $c_a$ (resp. $r_b$ and $c_b$), equilibrium behavior will be accommodating if and only if players' impatience and costs to generating instability are high enough.
What if both players have a low enough cost to generating instability, or are patient enough, such that there exists a region where both players would like to generate instability if the other one were inactive?
There cannot be an accommodating equilibrium in that case: both player using their inactive benchmark strategy would lead to both generating strictly positive instability at some status quo, contradicting \hyref{prop:equilibrium-properties}{Proposition}.
The structure of deterrence equilibria comes from the fact that, when $\underline x_b^0<\underline x_a^0$, in any equilibrium, it must be the case that the stable region is reduced to a single point.

\begin{lemma}
    \label{lemma:singleton-stable-region-deterrence}
    If $\underline x_a^0>\underline x_b^0$, then at any equilibrium $\underline x =\overline x$.
\end{lemma}

\noindent
To see why this must be the case, note that if $\underline x<\overline x$, both players' value functions must equal the identity on $[\underline x,\overline x]$ as on that region no one is generating instability.
From \hyref{prop:individualcontrol-supersolution}{Corollary}, it must be that at least one of the players would like to instigate instability were their opponent passive throughout, i.e. $\underline x< \underline x_a^0$ or $\underline x_b^0<\overline x$. 
\hyref{lemma:singleton-stable-region-deterrence}{Lemma} then shows that if a player generates instability at a given status quo when their opponent is passive throughout (as in the inactive benchmark), then they would do the same in any equilibrium in which their opponent is passive around this state. 
Moreover, \hyref{lemma:singleton-stable-region-deterrence}{Lemma} and \hyref{prop:individualcontrol-supersolution}{Proposition} combined imply that if $\underline x_a^0>\underline x_b^0$, then $\overline x = \underline x \in [\underline x_b^0,\underline x_a^0]$.

The second key observation is that, at any equilibrium such that $\underline x_a^0>\overline x> \underline x_b^0$, the players' equilibrium strategies are vanishing abruptly at $\overline x$.
Player $A$'s value function is a viscosity solution $v_a$ that satisfies $F_a(x,v_a(x),v_a''(x))=0$ on $[0,\overline x]$, $B(0,v_a'(x))=0$, and $v_a(\overline x)=\overline x$.
Consequently, we will have that the left-derivative of the value function at $\overline x$ is strictly smaller than one, as $v_a\leq v_a^0$ and
$v_{a,-}'(\overline x)\leq v_{a,-}^{0\prime}(\overline x)< v_{a,-}^{0\prime}(\underline x_a^0)\leq 1$, where the last inequality follows from the fact that $v_a^0$ is strictly convex on $(\overline x,\underline x_a^0)$.
As, owing to the regularity of our problem, we can derive, for $x \in (0,\overline x)$,
$$a^{*\prime}(x)\propto v_a'''(x)=r^2c\frac{v_a'(x)-1}{v_a''(x)}<0,$$
we find that $a^{*\prime}(x)\to -\infty$ as $x \uparrow \overline x$, given that the numerator is bounded away from zero and strictly negative $v_{a,-}'(\overline x)<1$, and the denominator is vanishing.

This second observation indicates that, at any equilibrium such that $\underline x_a^0>\overline x> \underline x_b^0$, players behave \emph{as if} $\overline x$ is an actively reflecting boundary.
Indeed, the fact that player $A$'s equilibrium strategy vanishes abruptly at $\overline x$ is reminiscent of how the optimal control in the inactive benchmark case is affected by an actively reflecting upper bound (\hyref{prop:properties-upper-reflection}{Proposition}): if $v_{a,-}^{0\prime}(\underline x_a^0)<1$, then ${a^{*,0\prime}_-}(\underline x_a^0)=-\infty$.

Combined, both these observations suggest a constructive method to characterize any equilibrium: 
take a candidate stable point $\overline x \in (\underline x_b^0,\underline x_a^0)$ and solve for the player $A$'s (resp. $B$'s) unique viscosity solution to the inactive benchmark problem on $[0,\overline x]$ (resp. $[\overline x, 1]$), as if reflection occurred at $\overline x$ instead of at 1 (resp. at 0).
Then, solve the HJB on the region in which the player is inactive taking the opponent's strategy as given with the appropriate boundary conditions, piece the two together, and verify that the resulting function is a viscosity solution to the original problem taking the opponent's strategy as given.
In particular, the resulting function needs not only to be continuous at the threshold, but it also cannot exhibit a convex kink at $\overline x$.

The observations above indicate that any equilibrium must conform with this construction, and thus it is pinned-down by the threshold $\overline x$.
The question of existence of an equilibrium can then be rephrased as follows: is there an state $\overline x \in[\underline x_b^0,\underline x_a^0]$ for which such a construction holds?
\hyref{thm:eqm-characterization}{Theorem} answers this question affirmatively and provides an exhaustive characterization: whenever $\underline x_b^0<\underline x_a^0$, 
\emph{every} $\overline x \in[\underline x_b^0,\underline x_a^0] \setminus \{0,1\}$ determines an equilibrium (when taken as the stable point in the construction above), and \emph{all} equilibria correspond to this construction for \emph{some} $\overline x \in[\underline x_b^0,\underline x_a^0] \setminus \{0,1\}$.

The proof verifies that any equilibrium needs to satisfy the construction laid out above, and that such a construction is successful in characterizing equilibrium viscosity solutions whenever $\overline x \in [\underline x_b^0,\underline x_a^0]\setminus \{0,1\}$.
Note that $\overline x$ can never be equal to 0 or 1 because, as proved in \hyref{thm:best-response-characterization}{Theorem}, $v_a(0),v_b(1)>0$: at extreme states, the player with nothing to lose generates strictly positive instability in equilibrium.

\begin{figure}
    \centering
    \makebox[\textwidth][c]{
    \begin{minipage}{1.0\linewidth}
        \begin{subfigure}{.5\linewidth}
            \includegraphics[width=1\linewidth]{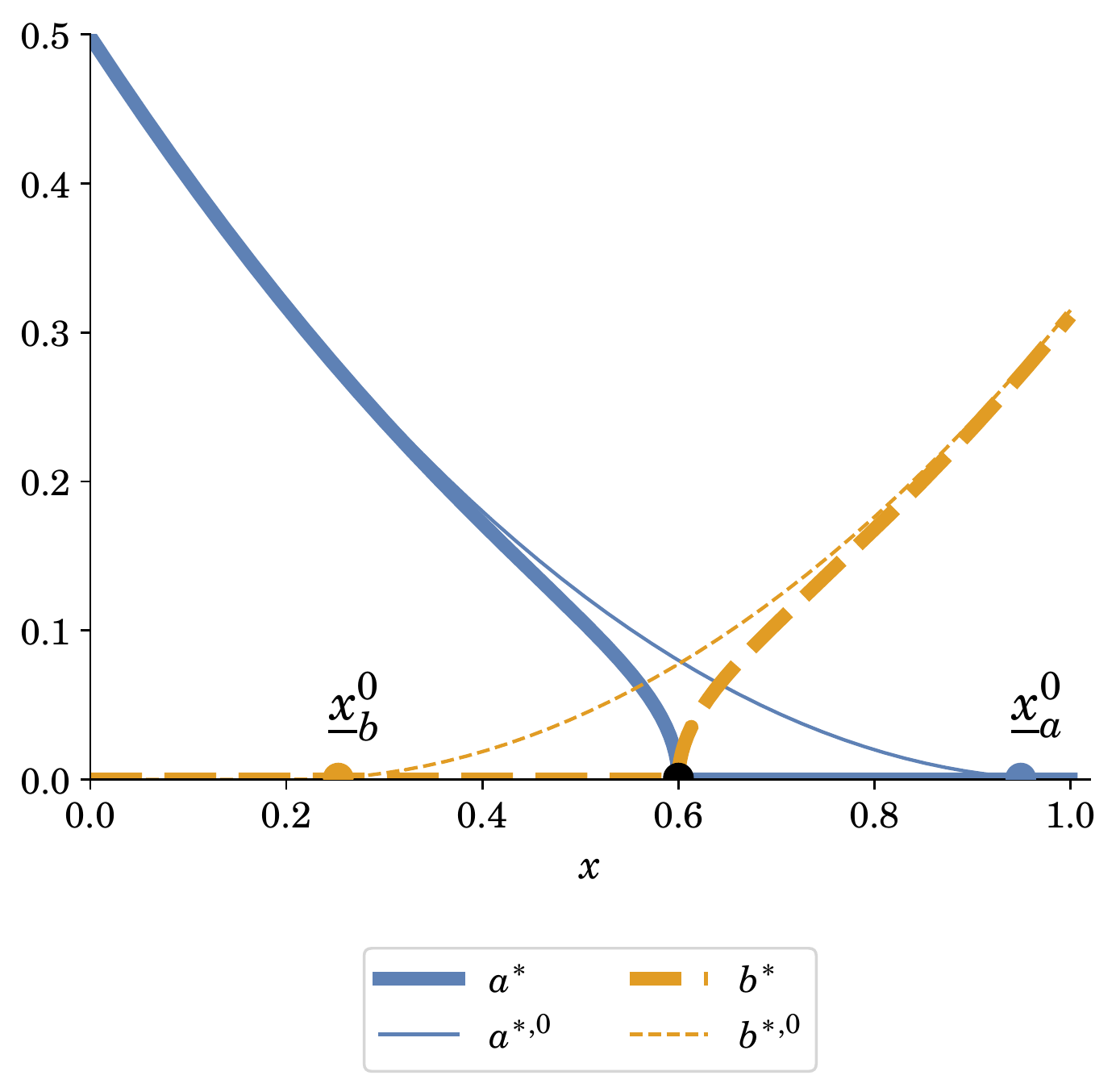}
            \caption{Equilibrium with $\underline x=\overline x =.6$}
            \label{fig:strat-DEqm:1}
        \end{subfigure}
        \begin{subfigure}{.5\linewidth}
            \includegraphics[width=1\linewidth]{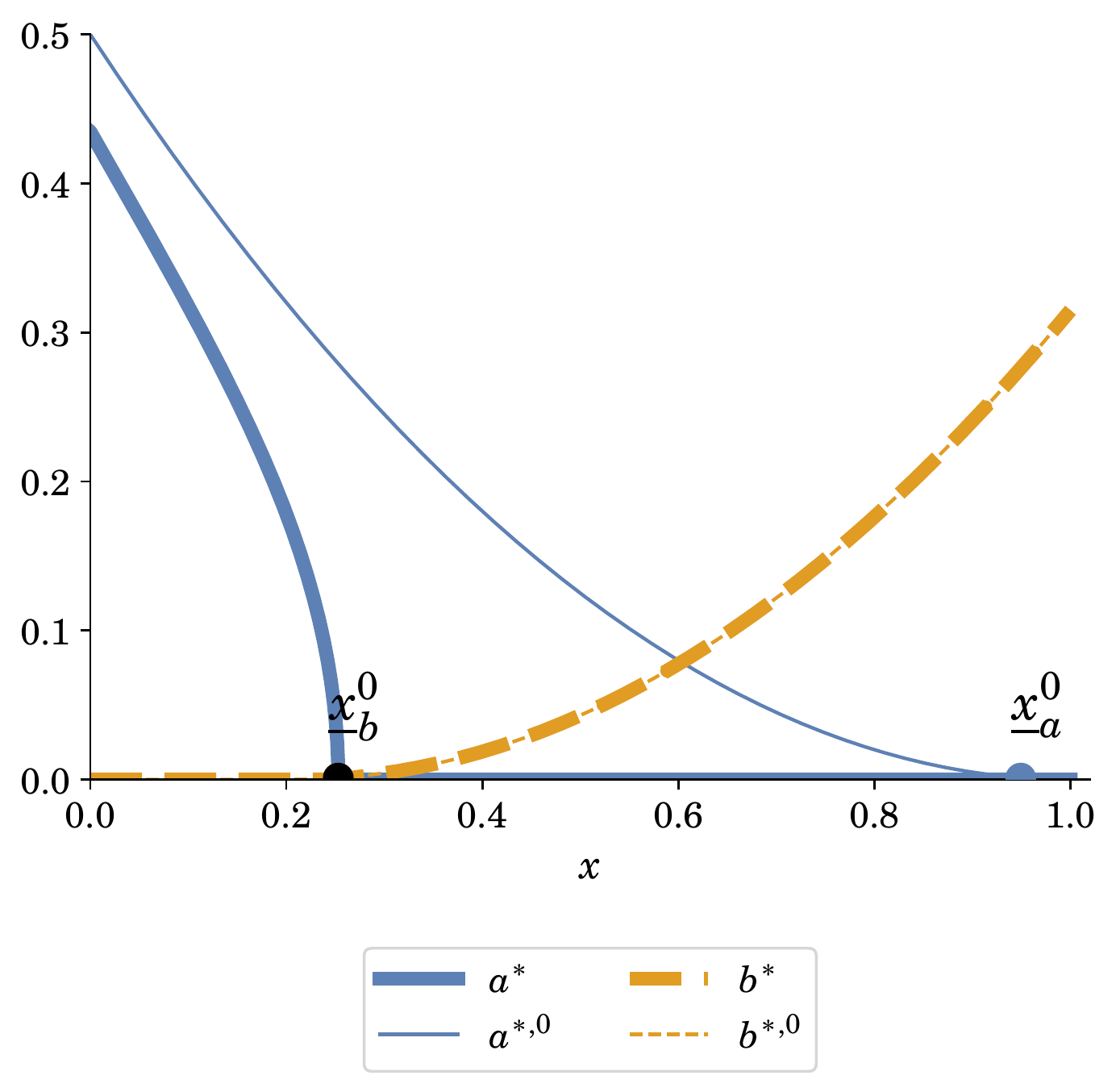}
            \caption{Equilibrium with $\underline x=\overline x =\underline x_b^0$}
            \label{fig:strat-DEqm:2}
        \end{subfigure}
    \end{minipage}
    }
    \caption{Equilibrium Strategies in a Deterrence Equilibrium}~\\
    \begin{minipage}{.95\linewidth}
        \footnotesize Note: Parameter values are $r_a=5$, $c_a=6$, $r_b=6$, and $c_b=15$.
    \end{minipage}
    \label{fig:strat-DEqm}
\end{figure}

In such cases, stability at equilibrium is sustained via deterrence: if their opponent were to not react, both players would like to destabilize the status quo in hope of an improvement of their situation at $\overline x$.
It is exactly because opponents would react and push back, and would do so with enough intensity, that $\overline x$ is a stable status quo.
This is related to the existence and interpretation of a concave kink in the value function.
There is a kink only in one very specific situation: a deterrence equilibrium, with a single stable status quo, supported by threats of high enough instability on both sides of it.
Indeed, if there is a kink at $\overline{x}$, then $b$ is strictly positive on $(\overline x,1]$ but zero at $\overline{x}$, which implies that $\overline{x}$ is an status quo at which neither player generates instability.

The fact that only concave kinks are possible can be interpreted as each player fighting back `hard enough' towards the stable status quo, so as to dissuade the other player from attempting to further improve their situation.\footnote{
    If there were a convex kink at $\overline x$, $v_a$ would be increasing faster to the right of $\overline{x}$ than to the left, making it profitable for player $A$ to strictly increase volatility in a way that pushes the process up and symmetrically for player $B$.
}
In a loose sense, it is the fact that player $B$ pushes back by abruptly increasing instability to the right of the stable status quo that renders it absorbing by making the slope of $v_a$ become discontinuously flatter. 
This deters player $A$ from taking action as it would be too costly to push the process beyond such a point, and again a symmetric argument holds for player $B$.

Equilibrium strategies in accommodating and deterrence equilibria exhibit meaningfully different properties: player $A$'s ($B$'s) equilibrium strategy is convex if $\underline x_a^0\leq \underline x$ ($\overline x\leq \underline x_b^0$) and convex-concave otherwise.
In an accommodating equilibrium (as in \hyref{fig:strat-AEqm:1}{Figure}), both players' equilibrium strategies are convex and instability vanishes smoothly.
In contrast, in a deterrence equilibrium (as in \hyref{fig:strat-DEqm:1}{Figure}), equilibrium strategies are convex-concave and have infinite slopes at the stable status quo, just as they do in the inactive benchmark when the upper bound becomes a binding constraint.
This again captures the constrained nature of a deterrence equilibrium: it is as if the other player is acting as a reflecting barrier at the stable status quo.
In \hyref{fig:strat-DEqm:2}{Figure}, we also exhibit the case of a second deterrence equilibrium in which the unique stable status quo coincides with $\underline x_b^0$.
In such case, player $B$'s equilibrium strategy also coincides with their optimal control in the inactive benchmark.
With a reflecting boundary at $\overline x=\underline x_b^0$ and player $B$'s optimal control would not be affected --- explaining the convex structure of the control.
However, it curtails player $A$'s ambitions of reaching more favorable states ($\underline x_a^0>\overline x = \underline x_b^0$), giving rise to the convex-concave structure of their equilibrium instability strategy.

\subsection{Equilibrium Comparative Statics}

The characterization of equilibria highlights that the thresholds $\underline{x}_a^0,\underline{x}_b^0$ capture the maximal threat power of each player: 
there is no equilibrium with a stable status quo more favorable to player $A$ than $\max\{\underline{x}_a^0,\underline{x}_b^0\}$ and less favorable than $\min\{\underline{x}_a^0,\underline{x}_b^0\}$, while the opposite is true for player $B$.

This intrinsic dependence of equilibria on the inactive benchmark thresholds $\underline x_a^0, \underline x_b^0$ entails that we can directly harness the comparative statics on the individual decision problem (\hyref{prop:idm-comparative-statics}{Proposition}) to obtain comparative statics of equilibria with respect to the players' costs to instability, $c$, and their patience or discount rate, $r$.

When players are impatient enough and face high enough costs to instability we obtain a unique accommodating equilibrium with a thick region of stable states, $[\underline x_a^0,\underline x_b^0]$, with $\underline x_a^0\leq \underline x_b^0$. 
Fixing $r_b,c_b$ (hence $\underline x_b^0$), as $r_a$ or $c_a$ decrease, $\underline x_a^0$ increases, and 
the set of stable states shrinks, 
as depicted in \hyref{fig:strat-AEqm:2}{Figure}: 
player $A$ will now find it worthwhile to generate instability at states that were previously stable.
With enough patience and costs low enough ($r_a,c_a$ small enough), we eventually obtain $\underline x_a^0 > \underline x_b^0$, and transition into deterrence equilibria. 
All equilibria have a unique stable stable state $\overline x \in [\underline x_b^0,\underline x_a^0] \setminus \{0,1\}$.
Hence, as $r_a,c_a$ further decrease, $\underline x_a^0$ increases and the set of equilibria expands. 
Let $\leq_{SSO}$ denote the strong set order. 
The following corollary summarizes the results: 

\begin{corol}
    \label{corol:comparative-statics-equilibrium}
    Fix $r_b,c_b>0$. 
    \begin{enumerate}[label=(\roman*)]
        \item There exists a unique $\theta>0$ such that $r_a^2 c_a \leq \,(<)\,  \theta$ if and only if $\underline x_a^0 \geq\, (>)\, \underline x_b^0$.
        \item Let $(r_a,c_a),(\tilde r_a,\tilde c_a)$  and, given $(r_b,c_b)$, denote by $\mathcal S$ and $\tilde{\mathcal S}$ 
        the sets of stable states associated with the respective equilibria.
        If $r_a^2c_a \leq \,(<)\,\tilde r_a^2 \tilde c_a$, then 
        equilibrium stable states increase in the strong set order, 
        $\mathcal{S} \leq_{SSO} \,(<_{SSO})\,\tilde{\mathcal{S}}$.
        Moreover, 
        $\tilde{\mathcal{S}} \supseteq \, (\supset)\, \mathcal{S}$ if $\theta \leq r_a^2c_a$, and 
        $\mathcal{S} \subseteq \, (\subset)\, \tilde{ \mathcal{S}}$ if $\tilde r_a^2\tilde c_a\leq \theta$.
    \end{enumerate}
\end{corol}

\noindent The result holds symmetrically if we fix $r_a,c_a$ and vary player $B$'s parameters. 

Since if a player is more patient (lower $r$) or faces lower costs to instability ($c$), equilibrium stable states shift in the strong set order, this suggests the possibility of obtaining comparative statics also with respect to equilibrium payoffs.
The comparison is straightforward across accommodating equilibria, but there is a subtlety when considering deterrence equilibria. 
Decreasing $c$ will still make a player everywhere better off since it shifts the value function up in their instability region, and leaves it unchanged in the passive region. 
Decreasing $r$, however, now has an ambiguous effect: 
fixing a given equilibrium (i.e. a stable point) it makes the player \emph{better off} in their instability region and \emph{worse off} in their passive region.
The multiplicity of equilibria further muddles the comparison, since it is possible to select different equilibria under the different parameters and have crossing equilibrium value functions.
The following proposition summarizes comparative statics of equilibrium payoffs in $c$ and $r$ respectively --- the proof is direct from comparison principle arguments:
\begin{prop}
    \label{prop:comparative-statics-welfare-cost}
    Fix $r_b,c_b$. 
    Let $\tilde v_a,v_a$ and $\underline {\tilde x}, \underline x$ be player $A$'s value function and equilibrium thresholds associated with equilibria given $\tilde r_a, \tilde c_a$ and $r_a, c_a$, respectively, such that $(\tilde r_a,\tilde c_a)\leq (r_a,c_a)$ and $\underline {\tilde x}\geq \underline x$.
    Then $\tilde v_a \geq v_a$ on $[0, \underline {\tilde x}]$.
    If, furthermore, $\tilde r_a = r_a$ or $\underline {\tilde x}<x_b^0$, then $\tilde v_a \geq v_a$ on $[0,1]$.
\end{prop}

\section{Equilibrium Dynamics}
\label{sec:dynamics}
What is the effect of players using strategically generating uncertainty on the dynamics of instability and the evolution of the status quo?
The precise characterization of equilibrium in the previous section can be used to answer these questions directly.

A salient characteristic of our model is that all equilibria (of accommodating or deterrence type) display a form of path dependency.
Consider an arbitrary equilibrium with thresholds $\underline{x},\overline{x}$ partitioning the state space $[0,1]$, and denote $X_0$ the initial point of the process.
If we start at a stable state, $X_0 \in [\underline x, \overline x]$, this will remain the status quo forever since no player generates any instability.
Moreover, if the process starts in say $A$'s instability region $[0,\underline x]$, it will also remain in this region --- and similarly for $B$'s instability region $[\overline x, 1]$.
This comes from continuity of the process and the fact that the outer boundary ($0$ or $1$ respectively) is reflecting while the inner boundary ($\underline x$ or $\overline x$ respectively) is absorbing in equilibrium.
This implies that whichever player starts off as most disadvantaged, will in equilibrium remain so forever, and can at most hope to reach their least preferred stable status quo.

Does the process converge in the long run towards a stable status quo?
Or does instability perpetuate if we start in an instability region? 
Given the previous discussion, if there is (probabilistic) convergence from a player's instability region, it will be towards their least preferred stable status quo.
The next proposition confirms this intuition.

\begin{prop}
    \label{prop:eqm-convergence}
    Let $X_t$ be the process associated to equilibrium strategies $a^*,b^*$, and denote $\underline x, \overline x$ the corresponding equilibrium thresholds.
    Then, (i) if $X_0 \in [\underline x, \overline x]$, $X_t = X_0$ for all $t$; and (ii) if $X_0< \underline x$ (resp. $>\overline x$), $X_t$ converges almost surely to $\underline x$ (resp. $\overline x$).
\end{prop}

\noindent For the case $X_0 \in [\underline x,\overline x]$, the proof of \hyref{prop:eqm-convergence}{Proposition} is trivial given that the process is degenerate and there is no instability.
For $X_0< \underline x$, we can use a constructive approach to show that $X_t$ is a submartingale.
Indeed, construct the process $Y_t$ defined by $dY_t:= \sqrt{2a^*(|X_t|)}dB_t$.
Because of the structure of $a^*$, $Y_t$ has absorbing boundaries on $[-\underline x, \underline x]$, and therefore we can verify that it is a martingale using boundedness of $a^*$ and the optional stopping theorem.
By using pathwise uniqueness of the solution $X_t$, we can argue that $Y_t=|X_t|$, that is, $Y_t$ is the mirror image of $X_t$ without the reflection at $0$ --- this is done by fixing a Brownian path $B_t(\omega)$, which uniquely pins down $X_t(\omega)$ by pathwise uniqueness.
Then, we argue that $Y_t$ and $X_t$ can only cross the origin at the same time and must be either identical or mirrored between two hitting times of zero (since they have the same increments).
Since the absolute value is a convex function, we conclude that $X_t$ is a submartingale, and by the martingale convergence theorem it must converge almost surely.
We can then prove that $X_t$ converges to $\underline x$ a.s. by contradiction, since convergence to any $x<\underline x$ would only be sustainable under a measure zero trajectory for the Brownian motion.
The argument for $X_0 > \overline x$ is symmetric using a similar construction around $1$.

\hyref{prop:eqm-convergence}{Proposition} entails instability is decreasing in the long run.
As players approach a stable status quo, whichever player is generating instability becomes more conservative --- a consequence of the properties of best responses.
Therefore, in the long run stability prevails.

\section{Discussion}
\label{sec:discussion}

We now discuss a number of variations on our model.

\noindent \textbf{Exogenous instability.} 
To clearly identify the strategic incentives to generate instability, we focused on the case in which any instability is endogenous. 
Given that our best-response characterization allows for arbitrary strategies by the opponent, and as these correspond to continuous exogenous state-contingent volatility structures, all in \hyref{sec:controlproblem:diff-characterization}{subsections} and \ref{sec:controlproblem:properties-br} holds identically when allowing for exogenous instability sources (independent from $\alpha_t$ and $\beta_t$, conditional on $X_t$). 
Considering a fixed exogenous level of instability $\sigma>0$, such that $\displaystyle dX_t = \sqrt{\alpha_t+\beta_t+\sigma^2}dB_t-dK_t$, an equilibrium exists.\footnote{
    It is easy to show that the unique viscosity solution $v_a$ given an arbitrary continuous $b$ is now thrice-continuously differentiable, and that the first three derivatives are bounded. 
    Existence of an equilibrium then follows by an application of Arzel\`{a}–Ascoli theorem and Schauder's fixed point theorem.
} 
However, while at any equilibrium there is a unique state $\overline x \in (0,1)$ at which $a^*(\overline x)=b^*(\overline x)=0$, the modified model (mechanically) exhibits perpetual instability.
There is no longer convergence to a stable status quo: the state will eventually become too unfavorable for any given player.
Consequently, players will forever alternate in creating instability so as to seek (temporary) improvements over the status quo.

\noindent \textbf{Costs to instability.} 
While we relied on quadratic costs for expositional convenience, provided enough regularity,\footnote{
    In particular, costs need to be sufficiently smooth, strictly increasing and strictly convex on $\mathbb R_+$, with $0=c(0)=c'(0)$.
    Although it goes beyond the scope of this paper to characterize its limits, the proof strategy to (a version of) \hyref{thm:HJBBC-controlproblem}{Theorem} extends given enough regularity on the cost function.
} 
results generalize to smooth, strictly convex costs to instability.  
In particular, the proofs for the threshold structure of best responses (and other properties in \hyref{thm:best-response-characterization}{Theorem}), monotonicity, and equilibrium characterization can be adjusted to accommodate general cost structures. 
The HJB equation would be given by
\[ r v_a(x)-\sup_{a \in \mathbb R_+}\left\{rx - r c(a)+(a+b(x))v''_a(x)\right\}.\]
Moreover, as $r c'(a^*(x))=v_a''(x)_+$, whenever $a^*(x)>0$ we would then have
\[ r(v_a(x)-x)-b(x)r c'(a^*(x))=r c'(a^*(x))a^*(x)-rc(a^*(x)),\]
from which one can obtain that monotonicity of $b$ implies monotonicity of $a^*$.\footnote{
    We thank Yu Fu Wong for having pointed out that monotonicity would extend for general cost structures in the individual decision-making case --- corresponding to our inactive benchmark with $b\equiv 0$. 
}

\noindent \textbf{State space.} A substantive assumption in our model is that the state space lies on a closed interval, as it is the option value provided by the lower bound that induces players to generate strictly positive instability. 
Absent a lower bound on the state space, players would have no desire to generate instability unless they were not risk-neutral. 
A similarly conclusion would hold if the boundaries were absorbing rather that reflecting.

\noindent \textbf{Terminal payoff.} Finally, we consider the case of having a terminal payoff, whereby the instead of accruing a flow benefit, players accrue that payoff only when the both players generate no instability. 
This can be seen as an extreme form of conflict, as creating instability fully deprives the opponent of any flow benefit. 
Immediately, for a given strategy of the opponent $b$, one can see that player $A$'s optimal control would need to satisfy $a(x)=0$ whenever $b(y)>0$ for any $y\geq x$, and thus we would have decoupling for any best response. 
Heuristically, in an inactive benchmark ($b\equiv 0$) one would expect player $A$'s value function to solve 
\[r v_a(x)=\max\left\{r x,\frac{1}{2rc} {[v_a''(x)_+]}^2\right\}\quad \text{ on } (0,1)\]
under boundary conditions $v_a'(0)=0,\quad v_a(1)=1$, with the control being given by $\displaystyle a^{*,0}(x)=\frac{1}{rc}v''_a(x)_+$. 
Differently from our model, we note that in such case the instability would be increasing rather than decreasing in the inactive benchmark.
Such a result is reminiscent of \citepos{GulPesendorfer2012REStud} and \citepos{Gieczewski2020WP}.\footnote{
    Also \citet{MoscariniSmith2001Ecta} who study not conflict but learning.
} 
Focusing on monotone strategies, a construction of accommodating equilibria with a region of stable states given by $[x_a^0,x_b^0]$ (if $x_a^0<x_b^0$), and of deterrence equilibria with a unique stable state $\overline x \in [x_b^0,x_a^0]$ (if otherwise) would be immediate,\footnote{
    Note that at a deterrence equilibrium, no player has an incentive to generate instability at $\overline x$ as this would lead to permanent instability and thus a null benefit to instability to both players.
} with instability greatest at states nearing the region of stable states.

\vspace*{2em}
Our model's novel approach to the mechanics of instability and its strategic importance in situations of conflict opens several paths for future investigation.
It demonstrates that the possible endogeneity of instability generates non-trivial dynamics that should be further investigated, notably to better understand the interaction of various conflict mechanisms in richer environments and their applications to concrete situations of conflict, bargaining, and related settings.

\renewcommand\bibname{~}
\setlength\bibhang{0pt}
\setlength{\bibsep}{0em plus 0ex}
\section{References} ~\\[-85pt]

\bibliographystyle{econ-aea}
{
    ~\\[-50pt]
    \bibliography{references}
}

\phantomsection
\addcontentsline{toc}{section}{Appendices}

\setcounter{section}{0}
\renewcommand{\thesection}{Appendix \Alph{section}}
\renewcommand{\thesubsection}{\Alph{section}.\arabic{subsection}}

\setlength{\parindent}{0in}
\section{Preliminaries}
\label{sec:appendix:preliminaries}
\subsection{Stochastic Differential Equations with Reflection}

 Consider our equation of interest, for a given $a,b$ continuous measurable functions:
 \[
 dX_t = \sqrt{2 \bigl(a(X_t) + b(X_t) \bigr) } dB_t - dK_t
 \]
With say $X_t, K_t$ solve the reflection problem on $\mathcal{O}:=(0,1)$ if they are the continuous $\mathcal{F}_t$-adapted processes such that (i) $dX_t = \sqrt{2 \bigl(\alpha_t + b(X_t) \bigr) } dB_t - dK_t$, (ii) $X_t \in [0,1]$ a.s., and (iii) $K_t$ is non-decreasing, its total variation $|K|_t = \int_0^t \mathds{1}_{X_t \in \{0,1\}} d|K|_s$, and $K_t = \int_0^t n(X_s) d|K|_s$, where $n(\cdot)$ denotes the unit outward normal vector to $\mathcal{O}$, that is, $n(1)=1$, $n(0)=-1$.
    
\noindent $K_t$ is the local time of the process at the boundary --- it minimally pushes $X_t$ back inside of the domain (towards the inner normal) if it hits the boundary by compensating the variations that would make $X_t$ exit the domain. \citet{LionsSznitman1984CommPureApplMath} show that such processes are uniquely defined in much more general reflecting domains, essentially under assumptions guaranteeing that the stochastic differential equation (SDE) without reflection has a strongly (pathwise) unique solution.\footnote{
    In general, this is not directly applicable to our equation. 
    It is well known since the seminal paper of \citet{YamadaWatanabe1971JMKU} that pathwise uniqueness of solutions to SDEs of the form $dX_t = \sigma(X_t) dB_t$ is difficult to guarantee beyond the general condition that $\sigma$ is H\"{o}lder continuous with coefficient at least $1/2$. This condition clearly does not hold for general $a,b$ continuous in our model. 
    However, subsequent work has improved on the H\"{o}lder-$1/2$ condition for specific cases. 
    For our case, the presence of the reflection helps guarantee existence and pathwise uniqueness although it might actually not hold for the unbounded domain. 
    In particular, \citet{BassChen2005IJS} proved that under mild regularity condition, the one-sided reflection problem has a pathwise unique for a $\alpha$-H\"{o}lder diffusion coefficient, $\alpha \in (0,1/2)$.
    \citet{BassBurdzyChen2007AP} extends and provides a different proof of the result. 
    Their proof strategy for the one-sided reflection essentially covers our case of interest and easily extends to having a second reflecting barrier: this guarantees the pathwise-uniqueness of a solution to our equation with one-sided reflection at zero.
    We can then complete the proof by using the analytical apparatus of \citet{LionsSznitman1984CommPureApplMath} or the original approach by \citet{Skorokhod1961TPA} to prove existence and pathwise-uniqueness with the second reflecting barrier given pathwise-uniqueness of the one-sided reflecting process.
}

\subsection{Test functions and Second-order Semijets}
We recall the definition of second-order semijets. 
The second-order subjet of $v$ at $x_0 \in (0,1)$ is denoted by $J^{2,-}_{[0,1]}v(x_0) \subset \mathbb{R}^2$ and defined as:
\begin{align*}
    (p,M) \in J^{2,-}_{[0,1]}v(x_0) \Longleftrightarrow v(x) \geq v(x_0) + p(x-x_0) + \frac{1}{2} M (x-x_0)^2 + o(|x-x_0|^2) \quad \text{ as } x \rightarrow x_0
\end{align*}
Because the bounds play a special role, when $x_0 \in \{0,1\}$ $x$ can only converge to $x_0$ from one side. 
Following \cite{CrandallIshiiLions1992BullAMS}, we consider the closure of the subjet $\overline{J}^{2,-}_{[0,1]}v(x)$ in to properly define the viscosity characterization (at the boundary and points of non-differentiability).

To relate the subjet with our definition of viscosity solutions in terms of test functions, we recall a classical result:
$(p,M) \in {J}^{2,-}_{[0,1]}v(x_0)$ if and only if there exists a $\mathcal{C}^2$ function $\phi$ such that $x_0$ is a local maximum of $v-\phi$ and $\phi'(x_0)=p$, $\phi''(x_0)=M$. 
It is without loss to require the maximum to be global and to impose $\phi(x_0)=v(x_0)$. 
In other words, the subjet contains the first- and second-order derivative values that are admissible for a smooth function $\phi$ that lies everywhere strictly \emph{below} $v$ (hence the \emph{sub}jet term) and equals $v$ at $x_0$. 
This captures all the relevant differential information on $v$ and can indeed be interpreted as a notion of differentiability for non-differentiable functions. 
The superjet is defined symmetrically, but considering a convex quadratic approximation (or a smooth test function) from above. 
We denote it  by $J^{2,+}_{[0,1]}v(x_0) \subset \mathbb{R}^2$ and it is defined as,
\begin{align*}
    (p,M) \in J^{2,+}_{[0,1]}v(x_0) \Longleftrightarrow v(x) \leq v(x_0) + p(x-x_0) + \frac{1}{2} M (x-x_0)^2 + o((x-x_0)^2) \quad \text{ as } x \rightarrow x_0
\end{align*}
with $\overline  J^{2,+}_{[0,1]}v(x_0)$ denoting the closure of the superjet.
The analogue result holds for test functions: $(p,M) \in \overline{J}^{2,+}_{[0,1]}v(x_0)$ if and only if there exists a $\mathcal{C}^2$ function $\phi$ such that $x_0$ is a local (wlog global) minimum of $v-\phi$ and $\phi'(x_0)=p$, $\phi''(x_0)=M$ (wlog $\phi(x_0)=v(x_0)$).

We alternate between the (equivalent) formulation of viscosity properties in terms of test functions and semijets, in order to choose the most convenient and intuitive approach.

\section{Omitted Proofs}
\label{sec:appendix:proofs}

\subsection{Proof of \hyref{thm:HJBBC-controlproblem}{Theorem} (Viscosity Characterization: Existence and Uniqueness)}
\label{sec:appendix:proofs-controlproblem}

\subsubsection{Proof of \hyref{prop:optimality-controlproblem}{Proposition} (Viscosity Characterization)}

    Recall the control problem:
    \begin{align*}
           v(x)  = \sup_{ \alpha \in \mathcal{A} } \; & \mathbb{E} \left[  \int_0^\infty e^{-rt} f(X_t,\alpha_t) dt \right]
           \quad 
            & \text{s.t. } dX_t = \sqrt{2 \bigl(\alpha_t + b(X_t) \bigr) } dB_t - n(X_t) dK_t
    \end{align*}
    where $f(x,a) = x - c\frac{a^2}{2}$.

    The proof of the viscosity characterization of the solution is standard and relies on applying the dynamic programming principle (DPP) and Ito's formula --- nonetheless, we could not find a derivation that exactly matches all of our assumptions, so  we provide a direct derivation following usual steps.\footnote{
        The closest result can be found in \citet{Lions1986Proc}, with more regularity assumptions adapted for a more general setting. 
        The following proof closely follows the approach in \citet[Section 4.3]{Pham2009Book}.
    } 
    Our specific setup allows us to greatly relax regularity assumptions on model primitives --- in particular, it is sufficient that $b$ is continuous as long as we can guarantee pathwise uniqueness for the SDE.

    We appeal to the following version of the DPP \citep[see e.g.][Section 3.3]{Pham2009Book} consisting of two results: (1) 
    for all $\alpha \in \mathcal{A}$, for all stopping time $\tau$, 
    $\displaystyle
    v(x) \geq \mathbb{E} \bigg[  \int_0^\tau e^{-rt} f(X_t^x,\alpha_t) dt + e^{-r\tau} v(X_\tau^x)\bigg]
    $;
    and (2) for all $\epsilon > 0$, there exists $\alpha \in \mathcal{A}$ such that for all stopping time $\tau$,
    $\displaystyle
    v(x) - \epsilon \leq \mathbb{E} \bigg[  \int_0^\tau e^{-rt} f(X_t^x,\alpha_t) dt + e^{-r\tau} v(X_\tau^x)\bigg]
    $,
    where we use the notation $X_t^x$ to denote the value at $t$ of the process following the $dX_t = \sqrt{2 \alpha_t + b(X_t) } dB_t - n(X_t) dK_t$ and starting from $X_0=x$.

    We first prove that $v$ is a supersolution to  \hyperref[eq:HJB]{\color{black}(RP)}. 
    Consider $x_0 \in [0,1]$ and $\varphi \in \mathcal{C}^2([0,1])$ such that $x_0$ is a global minimum of $v_*-\varphi$ and without loss $\phi(x_0)=v_*(x_0)$, where $v_*$ denotes the lower-semicontinuous (l.s.c.) envelope of $v$. 
    By definition, $\exists\{x_n\}_n$ such that $x_n \rightarrow x_0$ and $v(x_n) \rightarrow v_*(x_0)$ as $n$ goes to infinity. 
    By continuity of $\phi$, $\gamma_n:= v(x_n)-\phi(x_n) \rightarrow v_*(x_0)-\phi(x_0) = 0$. 
    Define $h_n$ to be any strictly positive sequence such that $h_n \rightarrow 0$ and $\gamma_n/h_n \rightarrow 0$ as $n$ goes to infinity. 
    Fix an arbitrary $\eta>0$ and define the stopping time $\tau_n:=\inf \{ t \geq 0, |X_t^{x_n} - x_n | > \eta \}$ (i.e. the first exit time of the process starting at $x_n$ from a ball of size $\eta$). 
    In turn define the stopping time $\theta_n:= \tau_n \wedge h_n$.

    Apply the DPP at $x_n$ using an arbitrary constant strategy $\alpha_t \equiv a$ and stopping time $\theta_n$:
    \begin{align*}
        v(x_n) \geq \mathbb{E} \left[  \int_0^{\theta_n} e^{-rt} f(X_t^{x_n},a) dt + e^{-r\theta_n} v(X_{\theta_n}^{x_n})\right].
    \end{align*}
    Since $x_0$ is a global minimum of $v_*-\phi$, $v(x) \geq v_*(x) \geq \phi(x)$ for all $x \in [0,1]$, and by construction $v(x_n) = \phi(x_n) + \gamma_n$, hence:
    \begin{align*}
    \varphi(x_n) + \gamma_n \geq \mathbb{E} \left[  \int_0^{\theta_n} e^{-rt} f(X_t^{x_n},a) dt + e^{-r\theta_n} \varphi(X_{\theta_n}^{x_n})\right].
    \end{align*}
    Applying Ito's formula at $\theta_n,x_n$ and rearranging yields:
    \begin{align*}
        \gamma_n &\geq \mathbb{E} \biggl[ \int_0^{\theta_n} e^{-rt} \left( f(X_t^{x_n},a) + (a+b(X_t^{x_n}))\varphi''(X_t^{x_n}) - r \varphi(X_t^{x_n}) \right) dt 
        \\
        &\quad - \int_0^{\theta_n} e^{-rt} \varphi'(X_t^{x_n}) n(X_t^{x_n}) dK_t + \int_0^{\theta_n} e^{-rt} \varphi'(X_t^{x_n})\sqrt{2(a + b(X_t^{x_n}))} dB_t \biggr].
    \end{align*}
    The integrand in the last term $\int_0^{\theta_n} \varphi'(X_t^{x_n})(a + b(X_t^{x_n})) dB_t$ is bounded, so the expectation is equal to zero. 
    Rearranging and diving by $h_n$ yields:
    \begin{align*}
        \frac{\gamma_n}{h_n} + \mathbb{E} \Biggl[ \frac{1}{h_n} \int_0^{\theta_n}  e^{-rt} \left( r \varphi(X_t^{x_n}) - f(X_t^{x_n},a) - (a+b(X_t^{x_n})) \varphi''(X_t^{x_n}) \right) dt \Biggr]
        +  \mathbb{E} \Biggl[ \frac{1}{h_n} \int_0^{\theta_n}\varphi'(X_t^{x_n}) n(X_t^{x_n}) & dK_t \Biggr]  \geq 0.
    \end{align*}
    For $n$ high enough, $\theta_n = h_n$ a.s. by continuity a.s. of trajectories of $X_t$. 
    We use dominated convergence and the mean value theorem to get that, when $n$ goes to infinity,
    \begin{align*}
        r \varphi(x_0) - f(x_0,a) - (a+b(x_0)) \varphi''(x_0) + \varphi'(x_0) n(x_0) \mathds{1}_{x_0 \in \{0,1\}}  \geq 0,
    \end{align*}
    where the last term comes by definition given $dK_0 = 0$ if $x_0 \in (0,1)$. 
    Hence:
    \begin{align*}
        r \varphi(x_0)  - \sup_{a \geq 0} \biggl\{ f(x_0,a) + (a+b(x_0)) \varphi''(x_0) \biggr\} + \varphi'(x_0) n(x_0) \mathds{1}_{x_0 \in \{0,1\}}  \geq 0.
    \end{align*}
    This implies that for all $x \in (0,1)$, $r \varphi(x_0)  - \sup_{a \geq 0} \bigl\{ f(x_0,a) + (a+b(x_0)) \varphi''(x_0) \bigr\} \geq 0$; at the boundary for $x \in \{0,1\}$, either $r \varphi(x_0)  - \sup_{a \geq 0} \bigl\{ f(x_0,a) + (a+b(x_0)) \varphi''(x_0) \bigr\} \geq 0$ or $\varphi'(x_0) n(x_0) \geq 0$ (it cannot be that both are negative since their sum is nonnegative). 
    From which we directly conclude that $v$ is a supersolution to \hyperref[eq:HJB]{\color{black}(RP)}.

    We now prove that $v$ is a subsolution to \hyperref[eq:HJB]{\color{black}(RP)}. Consider $x_0 \in [0,1]$ and $\varphi \in \mathcal{C}^2([0,1])$ s.t. $x_0$ is a global maximum of $v^*-\varphi$ with $\varphi(x_0) = v^*(x_0)$, where $v^*$ denotes the upper-semicontinuous (u.s.c.) envelope of $v$. 
    Assume by contradiction that $v$ is not a subsolution of \hyperref[eq:HJB]{\color{black}(RP)}. 
    
    Since $\varphi'(x_0) n(x_0) \mathds{1}_{x \in \{0,1\}}$ is strictly positive on the boundary and zero away from it and $x_0 \mapsto r \varphi(x_0)  - \sup_{a \geq 0} \biggl\{ f(x_0,a) + (a+b(x_0)) \varphi''(x_0) \biggr\}$ is continuous, there exists $\epsilon > 0$ and $\eta > 0$ such that for all $x \in B(x_0,\eta) \cup [0,1]$,
    \[
        r \varphi(x)  - \sup_{a \geq 0} \biggl\{ f(x,a) + (a+b(x)) \varphi''(x) \biggr\} + \varphi'(x) n(x) \mathds{1}_{x \in \{0,1\} } \geq \epsilon.
    \]
    Then by definition of the u.s.c. envelope we can consider a sequence $x_n$ taking values in $B(x_0,\eta) \cup [0,1]$ such that $x_n \rightarrow x_0$ and $v(x_n) \rightarrow v^*(x_0)$ as $n$ goes to infinity. Just as before, we denote $\gamma_n:= v(x_n) - \varphi(x_n) \rightarrow 0$ and $h_m$ a strictly positive sequence such that $h_m \rightarrow 0$ and $\gamma_m / h_m \rightarrow 0$.

    Define the stopping times $\tau_n:= \inf \{ t \geq 0, |X^{x_n}_t-x_n| > \eta' \}$ for some $\eta'$ such that $0 < \eta' < \eta$ and $\theta_n:= \tau_n \wedge h_n$. By the second part of the DPP stated above applied to with $\epsilon h_n / 2$ and taking stopping time $\theta_n$, there exists $\alpha^n \in \mathcal{A}$ such that:
    \begin{align*}
        v(x_n) - \frac{\epsilon h_n}{2} \leq \mathbb{E} \left[  \int_0^{\theta_n} e^{-rt} f(X_t^{x_n},\alpha^n_t) dt + e^{-r\theta_n} v(X_{\theta_n}^{x_n})\right]
    \end{align*}
    Recall that by construction $v(x_n) = \varphi(x_n) + \gamma_n$ and $v^* \leq \varphi$, hence
    \begin{align*}
        \varphi(x_n) + \gamma_n - \frac{\epsilon h_n}{2} \leq \mathbb{E} \left[  \int_0^{\theta_n} e^{-rt} f(X_t^{x_n},\alpha^n_t) dt + e^{-r\theta_n} \varphi(X_{\theta_n}^{x_n})\right]
    \end{align*}
    
    Applying Ito's formula to $e^{-r \theta_n} \varphi(X_{\theta_n}^{x_n})$ and rearranging gives:
    
    \begin{align*}
        \gamma_n - \frac{\epsilon h_n}{2} \leq \mathbb{E} & \Biggl[ \int_0^{\theta_n} e^{-rt} \left\{ \left( - r \varphi(X_t^{x_n}) + f(X_t^{x_n},\alpha^n_t) + (\alpha_t^n + b(X_t^{x_n})) \varphi''(X_t^{x_n}) \right) dt - \varphi'(X_t^{x_n}) n(X_t^{x_n}) dK_t \right\} \Biggr]
        \\
        & - \mathbb{E} \Biggl[ \int_0^{\theta_n} e^{-rt} \varphi'(X_t^{x_n}){\sqrt{2(\alpha_t^n + b(X_t^{x_n}))}}  dB_t  \Biggr]
    \end{align*}
    
    As $b$ is continuous by assumption, $\left| \varphi'(X_t^{x_n}){\sqrt{2(\alpha_t^n + b(X_t^{x_n}))}} \right|$ is bounded (because $X_t$ is bounded by construction) and the last expectation term is zero. Simplifying and dividing by $h_n$:
    \begin{align*}
        \frac{\gamma_n}{h_n} - \frac{\epsilon}{2} + \mathbb{E} \Biggl[ \frac{1}{h_n} \int_0^{\theta_n} e^{-rt} \left\{ \left( r \varphi(X_t^{x_n}) - f(X_t^{x_n},\alpha^n_t) - (\alpha_t^n + b(X_t^{x_n})) \varphi''(X_t^{x_n}) \right) dt + \varphi'(X_t^{x_n}) n(X_t^{x_n}) dK_t \right\} \Biggr] \leq 0
    \end{align*}
    By construction the term inside the integral is always greater than $\epsilon$, hence we find:
    \begin{align*}
     \frac{\gamma_n}{h_n} + \epsilon \left( \frac{\mathbb{E}[\theta_n]}{h_n} - \frac{1}{2}  \right) \leq 0
    \end{align*}
    Since by construction $\frac{\mathbb{E}[\theta_n]}{h_n}$ converges to $1$ when $n$ goes to infinity ($h_n$ goes to zero), so we obtain a contradiction and this concludes the proof.

\subsubsection{Proof of \hyref{prop:existenceuniqueness-controlproblem}{Proposition} (Existence and Uniqueness in the Control Problem)}

    The proof of \hyref{prop:existenceuniqueness-controlproblem}{Proposition} relies on a standard strategy: we first prove a comparison principle for our problem (every supersolution is above every subsolution); we then establish existence using Perron's method.
    The combination of those two results gives uniqueness and continuity.

    We first outline the proof structure for the comparison principle before detailing its steps.\footnote{
        The proof strategy is similar in spirit to standard proofs in the literature  \citep[e.g.][]{CrandallIshiiLions1992BullAMS}, but, because of the presence of non-Lipschitz terms in the HJB equation, parts of the canonical approximation methods will fail. 
        Hence we to appeal to arguments that are specific to the problem (which would generally be quite ill-conditioned).
    } 
    Take an arbitrary supersolution $\overline{w}$ (l.s.c. without loss) and an arbitrary subsolution $\underline{w}$ (u.s.c. without loss), and assume towards a contradiction that $\sup_{x \in [0,1]} \underline{w}(x)-\overline{w}(x) > 0$. 
    Note the supremum is attained and denote by $x^*$ a point at which it is. 
    
    We first show the supremum cannot be attained inside the domain, i.e. $x^* \notin (0,1)$, using standard approximation techniques for viscosity solutions (dedoubling variables and Ishii's lemma). 
    
    We then consider $x^*=0$. 
    We show that $\overline{w}$ is non-increasing in some neighborhood to the right of $0$; furthermore if either $\underline{w}(0) > 0$ or $b(0)>0$, then $\underline{w}$ is non-decreasing in some neighborhood to the right of $0$.
    $\overline{w}(0)<0$ implies $b(0)>0$; so, if $\underline{w}(0)>\overline{w}(0)$ then either $\underline{w}(0) > 0$ or $b(0)>0$.
    Therefore by the previous point $\underline{w}$ is non-decreasing in some neighborhood to the right of $0$.
    This yields a contradiction because if $\underline{w}$ is non-decreasing and $\overline{w}$ is non-increasing in a neighborhood of $0$ to the right, the supremum cannot be attained at $0$. 
    
    Next, symmetrically consider $x^*=1$. 
    We show that $\overline{w}$ is non-decreasing in some neighborhood to the left of $1$; furthermore if either $\underline{w}(1) > 1$ or $b(1)>0$, then $\underline{w}$ is non-increasing in some neighborhood to the left of $0$.
    Since $\overline{w}(1)<1$ implies $b(1)>0$, if $\underline{w}(1)>\overline{w}(1)$, then either $\underline{w}(1)>1$ or $b(1)>0$.
    And by the previous point $\underline{w}$ is non-increasing in some neighborhood to the left of $1$ --- hence similarly the supremum cannot be attained at $1$.
    
    This gives a contradiction, so we conclude $\sup_{x \in [0,1]} \underline{w}(x)-\overline{w}(x) \leq 0$, implying $\underline{w} \leq \overline{w}$ for all $x$.

    \begin{proof}
        Consider $\overline{w}$ a supersolution to \hyperref[eq:HJB]{\color{black}(RP)} and $\underline{w}$ a subsolution to \hyperref[eq:HJB]{\color{black}(RP)}. 
        Without loss of generality, assume $\overline{w}$ to be l.s.c. and $\underline{w}$ to be u.s.c. --- the proof goes through the same way for the l.s.c. (resp. u.s.c.) envelope of $\overline{w}$ (resp. $\underline{w}$), in turn giving the same result since $\overline{w}(x) \geq \overline{w}_*(x) \geq \underline{w}^*(x) \geq \underline{w}(x)$.

        Assume by contradiction that $\sup_{x \in [0,1]} \underline{w}(x)-\overline{w}(x) > 0$. 
        This supremum is attained (since $\underline{w}-\overline{w}$ is u.s.c.) and we denote $x^*$ a point which attains it.
        
        We first show a maximum principle result: the supremum of $\underline{w}-\overline{w}$ cannot be attained in the interior of the domain, i.e. $x^* \in \{0,1\}$.
        Assume towards a contradiction that $x^*  \in (0,1)$. Define:
                \[
                    M_\alpha:= \sup_{x,y \in [0,1]} \underline{w}(x) - \overline{w}(y) - \frac{\alpha}{2} |x-y|^2
                \]
        this supremum is attained and we denote $(x_\alpha,y_\alpha)$ a point at which it is. 
        Clearly $M_\alpha \geq \underline{w}(x^*) - \overline{w}(x^*) > 0$.
        Furthermore $\lim_{\alpha \rightarrow \infty} \alpha |x_\alpha-y_\alpha|^2 = 0$ and $\lim_{\alpha \rightarrow \infty} M_\alpha = \underline{w}(x^*) - \overline{w}(x^*)$ (this is a general result, see for instance \citet[Lemma 3.1.]{CrandallIshiiLions1992BullAMS}).

        Let $f(x,y):=\underline{w}(x) - \overline{w}(y)$. 
        Using Ishii's Lemma \citep[Theorem 3.2.]{CrandallIshiiLions1992BullAMS}, we know that if $\psi \in \mathcal{C}^2([0,1]^2)$ is such that $(\hat{x},\hat{y})$ is a local maximum of $f-\psi$, then, for each $\epsilon > 0$ there exist $Y,X \in \mathbb{R}$ such that
        (1) 
        $(D_x \psi(\hat{x},\hat{y}),X) \in \overline{J}^{2,+}_{\mathcal{O}} \underline{w}(\hat{x})$, i.e. there exists $\underline{\varphi} \in \mathcal{C}^2$ such that $\hat{x}$ is a local minimum of $\underline{w}-\underline{\varphi}$ with $\underline{\varphi}'(\hat{x})=D_x \psi(\hat{x},\hat{y})$, and $\underline{\varphi} ''(\hat{x})=X$;
        and 
        (2) $(-D_y \psi(\hat{x},\hat{y}),Y) \in \overline{J}^{2,-}_{\mathcal{O}} \overline{w}(\hat{y})$, i.e. there exists $\overline{\varphi} \in \mathcal{C}^2$ such that $\hat{y}$ is a local maximum of $\overline{w}-\overline{\varphi}$ with $\overline{\varphi}'(\hat{y})=-D_y \psi(\hat{x},\hat{y})$, and $\overline{\varphi} ''(\hat{y})=Y$.
                 And we have
                    \[
                    - \left( 1 + ||D^2\psi(\hat{x},\hat{y}||) \right) I_2 \leq \begin{pmatrix} \underline{\varphi}''(\hat{x}) & 0 \\ 0 & -\overline{\varphi}''(\hat{y}) \end{pmatrix} \leq D^2\psi(\hat{x},\hat{y}) + \epsilon \left( D^2 \psi(\hat{x},\hat{y}) \right)^2.
                    \]
            
        Hence for any $\alpha>0$, we can take $\epsilon=1/\alpha$ and apply this result at $(x_\alpha,y_\alpha)$ with $\psi_\alpha(x,y):=\frac{\alpha}{2}|x-y|^2$.
        This implies there exists $\underline{\varphi}_\alpha,\overline{\varphi}_\alpha$ appropriate test functions for $\underline{w},\overline{w}$ respectively at $x_\alpha,y_\alpha$ such that $\underline{\varphi}''_\alpha(x_\alpha) \leq \overline{\varphi}''_\alpha(y_\alpha)$ for all $\alpha>0$. 
        Since $x^* \in (0,1)$, the supersolution and subsolution properties entail that for any $\alpha$: $F(x_\alpha,\underline{w}(x_{\alpha}),\underline{\varphi}_{\alpha}''(x_{\alpha})) \leq 0 \leq F(y_{\alpha},\overline{w}(y_{\alpha}),\overline{\varphi}_{\alpha}''(y_{\alpha}))$.
        Rearranging yields
                \[
                    r(\underline{w}(x_{\alpha})-\overline{w}(y_{\alpha})) - r (x_{\alpha} - y_{\alpha}) \leq b(x)  \left( \underline{\varphi}''_{\alpha}(x_{\alpha}) - \overline{\varphi}''_{\alpha}(y_{\alpha}) \right) + \frac{1}{2rc}  \left( [\underline{\varphi}''_{\alpha}(x_{\alpha})]^2 - [\overline{\varphi}''_{\alpha}(y_{\alpha})_+]^2 \right) \leq 0,
                \]
            which then implies
                   $ r(\underline{w}(x_{\alpha})-\overline{w}(y_{\alpha}) - \frac{{\alpha}}{2} |x_{\alpha}-y_{\alpha}|^2) - r (x_{\alpha} - y_{\alpha}) \leq 0.
                   $
        Taking the limit in the left hand side yields
                    $
                    \underline{w}(x^*) - \overline{w}(x^*) \leq 0.
                    $
            This contradicts our premise.
            Therefore if $\sup_{x \in [0,1]} \underline{w}(x)-\overline{w}(x) > 0$, the supremum can only be attained on the boundary, i.e. $x^* \in \{0,1\}$.

        Now consider the case $x^* = 0$. 
        We first prove that $\overline{w}$ is non-increasing in some right neighborhood of $0$. 
        By definition of the second-order subjet, it is sufficient to show that, for all $(p,M) \in \overline{J}^{2,-}_{[0,1]}\overline{w}(0)$, $p \leq 0$.
        
        Assume by contradiction there exists $(p,M) \in \overline{J}^{2,-}_{[0,1]}\overline{w}(0)$ such that $p > 0$. 
        Consider any $p'$ such that $0<p'<p$ and an arbitrary $M'>0$. 
        There must exist some neighborhood of $0$ (to the right) such that $px + \frac{1}{2} M x^2 \leq p'x + \frac{1}{2} M' x^2$ (the first-order terms dominate for $x$ small enough). 
        Therefore, as $x \rightarrow 0$,
        $\overline{w}(x) \geq \overline{w}(0) + px +  M x^2/2 + o(x^2) \geq \overline{w}(0) +  p'x + M' x^2/2 + o(x^2).$
        Hence $(p',M') \in J^{2,-}_{[0,1]}\overline{w}(0)$. Since this holds (close enough to zero) for $M'$ arbitrarily large, we get a contradiction since $B(0,p')<0$ and $F(0,\overline{w}(0),M')<0$ for $M'$ large enough.

        We claim that if either $\underline{w}(0)>0$ or $b(0)>0$, then $\underline{w}$ has to be non-decreasing in some neighborhood of $0$. 
        It is again sufficient to show that for all $(p,M) \in J^{2,+}_{[0,1]}\underline{w}(0)$, $p \geq 0$. 
        
        Assume by contradiction that there exists $(p,M) \in \overline{J}^{2,+}_{[0,1]}\underline{w}(0)$ with $p<0$. 
        Take any $p'$ such that $p < p' < 0$. For an arbitrary $M'<0$, there must exist some neighborhood of $0$ (to the right) such that $px + \frac{1}{2} M x^2 \leq p' x  + \frac{1}{2} M' x^2$ (the second-order terms vanish faster as $x$ goes to zero), hence, as $x \rightarrow 0$,
        $\underline{w}(x) \leq \underline{w}(0) + px +  M x^2/2 + o(x^2) \leq \underline{w}(0) + p'x + M' x^2/2 + o(x^2)$.
        This implies $(p',M') \in J^{2,+}_{[0,1]}\underline{w}(0)$. 
        Note $F(0,\underline{w}(0),M')=r\underline{w}(0)-b(0)M'>0$ when either $\underline{w}(0)>0$ or $b(0)>0$. 
        Hence this is a contradiction since $B(0,p) > 0$ and $F(0,\underline{w}(0),0)>0$.

        If $\overline{w}(0)<0$, it must be that $b(0)>0$.
        Indeed, if $b(0)=0$, then, by continuity, for all $\varepsilon>0$, $\exists x_\varepsilon>0$ such that $0 \leq b(x_\varepsilon) < \varepsilon$. 
        For any $\epsilon$, select arbitrarily $(p_\varepsilon,M_\varepsilon) \in \overline{J}^{2,-}_{[0,1]}\underline{w}(x_\varepsilon)$. 
        We have, for all $\varepsilon>0$,
            \begin{align*}
                0 \leq F(x_\varepsilon,\overline{w}(x_\varepsilon),M_\varepsilon) 
                = r\overline{w}(x_\varepsilon) - r x_\varepsilon - b(x_\varepsilon) M_\varepsilon - \frac{1}{2rc} {M_\varepsilon}_+^2
                \leq r \overline{w}(0) - b(x_\varepsilon) M_\varepsilon
            \end{align*}
        Since $b(x_\varepsilon)<\varepsilon$, this must imply that $M_\varepsilon < - \frac{r \overline{w}(0)}{\varepsilon}$.
        In other words, as we get close enough to zero the second-order terms in the subjets are bounded above by an arbitrarily negative constant. 
        This delivers a contradiction, since it would mean that $\overline{w}$ is locally bounded above by an arbitrarily concave paraboloid as we get closer to zero. 
        To make this point formal, define $M'_\varepsilon:= M_\varepsilon + \varepsilon$; from the previous point $(p_\varepsilon,M'_\varepsilon) \notin \overline{J}^{2,-}_{[0,1]}\underline{w}(x_\varepsilon)$, i.e., by definition, as $x \rightarrow x_\varepsilon$, 
        $\overline{w}(x) < \overline{w}(x_\varepsilon) + p_\varepsilon(x-x_\varepsilon) + M'_\varepsilon (x-x_\varepsilon)^2/2 + o((x-x_\varepsilon)^2)$.
        
        Defining $\varphi_\varepsilon(x):=\overline{w}(x_\varepsilon) + p_\varepsilon(x-x_\varepsilon) + \frac{1}{2} M'_\varepsilon (x-x_\varepsilon)^2$, $x_\varepsilon$ is \emph{not a local minimum} of $\overline{w}-\varphi_\varepsilon$. 
        But, by construction, since $M'_\varepsilon \rightarrow - \infty$ and $x_\varepsilon \rightarrow 0$ as $\varepsilon$ goes to zero, $\varphi_\varepsilon(x) \xrightarrow[\varepsilon \rightarrow 0]{} \mathds{1}_{x \neq 0} \times (- \infty)$, i.e., the function that has value $0$ at $0$, and negative infinity everywhere else, hence $\liminf_{x \rightarrow 0} \overline{w}(x) = - \infty < \overline{w}(0)$ contradicting that $\overline{w}$ is l.s.c.
        
        This entails that, if $\underline{w}(0)>\overline{w}(0)$, then $\underline{w}$ is non-decreasing in some neighborhood of $0$ to the right --- because either $\overline{w}(0) \geq 0$ which implies $\underline{w}(0)>0$ or $\overline{w}(0) < 0$ which implies $b(0)>0$.
        Therefore, we have that in some neighborhood of $0$ $\overline{w}$ is non-increasing and $\underline{w}$ is non-decreasing, which directly contradicts the fact that the supremum of $\underline{w}-\overline{w}$ is reached at $0$ and not in the interior.

        The only remaining possibility is $x^* = 1$. 
        The derivations are symmetrical to the previous case and we obtain that in some neighborhood of $1$ $\overline{w}$ is non-decreasing and $\underline{w}$ is non-increasing, which directly contradicts the fact that the supremum of $\underline{w}-\overline{w}$ is reached at $1$ and not in the interior.
        
        Putting those points together yields a contradiction. Therefore, we conclude that
        $
        \sup_{x \in [0,1]} \underline{w}(x)-\overline{w}(x) \leq 0,
        $
        which entails $\overline{w}(x) \geq \underline{w}(x)$ for all $x \in [0,1]$, and concludes the proof.
    \end{proof}

    \begin{lemma}[Existence --- Perron's Method]
        If the comparison principle holds for \hyperref[eq:HJB]{\color{black}(RP)}, and if there is a subsolution $\underline{w}$ and a supersolution $\overline{w}$ that satisfy the boundary conditions (in the viscosity sense), then
            $
            \hat{w}(x):= \sup \bigl\{ \; w(x): \underline{w} \leq w \leq \overline{w} \text{ and $w$ is a subsolution of \hyperref[eq:HJB]{\color{black}(RP)} } \bigr\}
            $
        is a solution of \hyperref[eq:HJB]{\color{black}(RP)}.
    \end{lemma}

    This is standard and can be directly applied from e.g. \cite{CrandallIshiiLions1992BullAMS}. 
    Furthermore, we can exhibit an explicit supersolution (take $\overline{w}(x):=1$ for all $x$) and an explicit subsolution (take  $\underline{w}(x):=0$ for all $x$), directly giving existence.

\subsection{Proof of \hyref{thm:best-response-characterization}{Theorem} (Best-Response Characterization)}

The proof of \hyref{thm:best-response-characterization}{Theorem} consists of the following intermediary results:

 \begin{prop}
    \label{prop:convex-concave}
    There are $\underline{x}_a,\overline{x}_a \in (0,1]$, $\underline{x}_a \leq \overline{x}_a$, such that
        (i) on $[0,\underline x_a)$, $v_a$ is convex and strictly above the identity;
        (ii) on $[\underline x_a,\overline x_a]$, $v_a$ is equal to the identity;
        and
        (iii) on $(\overline x_a,1]$, $v_a$ is concave and strictly below than the identity.
    Further, $v_a$ is increasing and $\forall x \in [0,1]$,\\ $\max\{\sup_{x \in [0,\overline x_a]} \underline \partial v_a(x), \sup_{x \in [\underline x_a,1]}\overline \partial v_a(x)\} \leq 1$.
\end{prop}
where $\underline \partial v_a$ and $\overline \partial v_a$ denote the sub- and supergradient of $v_a$ on $[0,\overline x_a]$ and $[0,\underline x_a]$;\footnote{
    That is, $\underline \partial v_a(x):=\{p \,\mid\, v_a(x')-v_a(x)\geq p(x'-x),\, \forall x' \in [0,\overline x_a]\}$ and 
    $\overline \partial v_a(x):=\{p \,\mid\, v_a(x')-v_a(x)\leq p(x'-x),\, \forall x' \in [\underline x_a,I]\}$.
} 
and
\begin{prop}
    \label{prop:C2}
        $v_a$ is of class $\mathcal{C}^2$ everywhere except possibly at $\overline{x}_a$ where it might not be differentiable. 
        Moreover, (i) $v_a'(0)=0$, and (ii) $v_a$ is not differentiable at $\overline{x}_a$ only if 
        (a) $\lim_{x \rightarrow \overline{x}_a^-} v'_a(x) \geq \lim_{x \rightarrow \overline{x}_a^+} v'_a(x)$, 
        (b) $b(\overline{x}_a)=0$, and 
        (c) if $\overline{x}_a<1$, then $b(1)>0$.
\end{prop}

\subsubsection{Proof of \hyref{prop:convex-concave}{Proposition} (Value Function is Convex-Concave)}
\begin{proof}
    By \hyref{prop:existenceuniqueness-controlproblem}{Proposition}, the unique viscosity solution $v$ is continuous.
    Let $X^>:=\{x \in [0,1]\,\mid\,v_a(x)>x\}$ and $X^<:=\{x \in [0,1]\,\mid\,v_a(x)<x\}$, $X^=:=[0,1]\setminus(X^>\cup X^<)$.
    As $v_a$ is a subsolution (resp. supersolution), $F_a(x,v_a(x),M):=r (v_a(x) -x)-b(x)M - \frac{1}{2r c}[M_+]^2$ and $b\geq 0$, and
    for any interval $I \subseteq X^>$ (resp. $I\subseteq X^<$) we have that $M>0$ (resp. $M<0$) for all $x \in I$ and all $(\psi,M) \in \overline J^{2,+}_{[0,1]}v_a(x)$ (resp. $\overline J^{2,-}_{[0,1]}v_a(x)$).
    Note that, on $X^=$, $v_a$ is linear.
    As, by \citet[Lemma 1]{AlvarezLasryLions1997JMathPuresAppl}, for any convex and open subset $I\subseteq X^>\cup X^=$ (resp. $I\subseteq X^<\cup X^=$), $v_a$ is convex (resp. concave) on $I$.
    
    We now show that for any element $x$ in an open interval $I \subseteq X^>$, its subgradient is such that $\max \underline \partial v_a(x)<1$.
    As $v_a$ is convex on $I$, its non-empty-, compact-, convex-valued, and non-decreasing.
    If $\max \underline \partial v_a(x)\geq 1$, then we have that $v_a(x')\geq v_a(x)+x'-x>x'$ for any $x'\in I$ such that $x'>x$.
    By continuity of $v_a$, $[x,1]\subseteq X^>$ and we obtain $v_a(1)>1$.
    However, as $v_a$ is a subsolution we must have that $0\geq \min\{F_a(1,v_a(1),M),B(1,p)\}=B(1,p)$ for any $(p,M) \in \overline J^{2,+}_{[0,1]}v_a(1)$.
    And, by convexity of $v_a$ on $[x,1]$ and the fact that $\max \underline \partial v_a(x)\geq 1$, we have that $(1,0)\in \overline J^{2,+}_{[0,1]}v_a(1)$, resulting in $B(1,p)=1>0$, a contradiction.
    An analogous argument holds to show that the supergradient of $v_a$ at any point $x$ of an open interval $I \subseteq X^<$ satisfies $\max \overline \partial v_a(x)<1$.
    
    The bound on the supergradient of $v_a$ implies that, if $x \in X^<$, it must be that $\forall x'\in [x,1]$, $v_a(x')<v_a(x)+x'-x<x'$ and thus $x' \in X^<$.
    Immediately, we obtain $\sup X^>\leq \inf X^\leq$.
    Hence $\exists \underline x_a,\overline x_a \in [0,1]$ such that $[0,\underline x_a)=X^>$, $[\underline x_a,\overline x_a]=X^=$, and $(\overline x_a,1]=X^>$, with $X^<$ and $X^>$ potentially empty.
    
    Next, we clarify that, in fact, $X^>,X^=\ne \emptyset$ (noting $X^>$ is an open set in $[0,1]$), by showing that $0 \in X^>$.
    Suppose instead $v_a(0)=0$ (and thus $X^\leq =[0,1]$, with $v_a$ concave on $[0,1]$).
    If there is some $x' \in [0,1]$ such that $v(x')>0$, let $p:=\frac{v_a(x')}{x'}>0$.
    As $v$ is concave, $v_a(x)=v_a(x)-v_a(0)\geq p(x-0)=p \cdot x>\frac{p}{2}(x+x^2)$ for all $x \in [0,x']$, and so $(\frac{p}{2},p)\in \overline J^{2,-}_{[0,1]}v_a(0)$ and $\max\{F_a(0,v_a(0),p),B(0,\frac{p}{2})\}<0$, contradicting that $v_a$ is a supersolution.
    If there is no such $x'$, then $v_a \equiv 0$ and $(0,-1)\in \overline J^{2,-}_{[0,1]}v_a(1/2)$, with $F(1/2,v_a(1/2),-1)=-1/2<0$, again contradicting $v_a$ is a supersolution.
    
    Our last step is to show $v_a$ is increasing.
    First note that, by convexity of $v_a$, $\max \overline \partial v_a(x)\leq \min \overline \partial v_a(x')$ for any $x,x' \in X^\geq$ such that $x'>x$.
    Suppose, for the purpose of contradiction, $\max \overline \partial v_a(0)<0$.
    This implies $\forall x \in (0, \overline x_a]$, $0>v_a(0)-v_a(x)$.
    Then, letting $p:=\frac{v_a(x)-v_a(0)}{x}<0$, we have $(p,0) \in \overline J^{2,+}_{[0,1]}v_a(0)$, which results in $\min\{F(0,v_a(0),0),B(0,p)\}>0$, a contradiction to $v_a$ being a subsolution.
    As, symmetrically on $[\underline x_a, 1]$, $v_a$ is concave and thus $\min \underline \partial v_a(x)\geq \max \underline \partial v_a(x')$ for $x,x' \in X^\leq$, it suffices to show $0 \in \underline \partial v_a(1)$.
    Suppose to the contrary that for some $x\in [\underline x_a,1)$, $1\geq v_a(x)>v_a(1)$.
    Then, $p:=\frac{v_a(1)-v_a(x)}{1-x}<0$ and $(p,0) \in \overline J^{2,-}_{[0,1]}v_a(1)$, implying $\max\{F_a(1,v_a(1),0),B(1,p)\}<0$, now a contradiction to $v_a$ being a supersolution.
\end{proof}

\subsubsection{Proof of \hyref{prop:C2}{Proposition} (Value Function is $\mathcal C^2 $, except possibly at a point)}
\begin{proof}
    \textbf{$v''$ exists a.e.:} From \hyref{prop:convex-concave}{Proposition}, $\exists \underline x_a,\overline x_a\in [0,1]$ such that $\underline x_a\leq \overline x_a$ and $v_a$ is convex on $[0,\overline x_a]$ and concave on $[\underline x_a,1]$. 
    By Alexandrov theorem, $v_a$ is twice differentiable a.e. on $[0,1]$, and so it has left- and right-derivatives everywhere, denoted by $v'_{a,-}$ and $v'_{a,+}$ respectively. 
    
    \textbf{No convex kinks:} Take any $x' \in [0,1]$. 
    Suppose by contradiction that $v'_{a,-}(x')<v'_{a,+}(x')$ and fix $p \in (v'_{a,-}(x'),v'_{a,+}(x'))$. 
    For any fixed $M>0$, $(p,M) \in \overline J^{2,-}_{[0,1]}v_a(x')$.\footnote{
        To see this, let $f(x):=v_a(x')+p(x-x')+\frac{1}{2}M{(x-x')}^2$, and note that $v_a-f\geq 0$ in a neighborhood of $x'$, therefore with $x'$ being a local minimum of $v_a-f$.
    } 
    However, for large enough $M$, $F_a(x',v_a(x'),M)=r(v_a(x')-x')-b(x')M-\frac{1}{2rc}M_+^2<0$, contradicting that $v_a$ is a supersolution. 
    
    \textbf{At most one concave kink at $\overline x$:} Now take any $x' \in [0,1]$. 
    Again suppose by contradiction $v'_{a,-}(x')>v'_{a,+}(x')$ and fix $p \in (v'_{a,+}(x'),v'_{a,-}(x'))$. 
    By a similar argument as before, for any fixed $M>0$, $(p,-M) \in \overline J^{2,+}_{[0,1]}v_a(x')$. 
    For $b(x')>0$ and large enough $M$, $F_a(x',v_a(x'),-M)=r(v(x')-x')+b(x')M>0$, which contradicts $v_a$ being a subsolution.
    For $b(x')=0$ and $x'\in(0,1)$, $F_a(x',v(x'),-M)=r(v_a(x')-x')\leq 0$.
    As $v_a$ is player $A$'s value function, whenever $b(x')=0$, the player can attain at least $v_a(x')\geq x'$ by setting the control to zero.
    Hence, we must have $v_a(x')=x'$. 
    As $v_a(x')=x'\Longleftrightarrow x' \in [\underline x_a, \overline x_a]$, we obtain $v'_a(x')=1$ for any $x' \in (\underline x_a, \overline x_a)$.
    On $[0,\overline x_a)$, $v_a$ is convex and $v'_{a,-}(x')\geq v'_{a,+}(x')$. 
    Thus, there are no concave kinks except possibly at $\overline x$ and only if $b(\overline x_a)=0$. 
    
    \textbf{Continuity of $v'_a$ on $[0,1]\setminus\{\overline x_a\}$:}
    On $[0,\overline x_a)$, $v'_a$ exists and is monotone as $v_a$ is convex (by \hyref{prop:convex-concave}{Proposition}). 
    As $v'_a$ is also differentiable a.e., it has the intermediate value property (by Darboux theorem), which, together with monotonicity, implies $v'_a$ is continuous on $[0,\overline x_a)$.
    A symmetric argument applies to $(\overline x_a,1]$.
    
    \textbf{Existence and continuity of $v''_a$ on $[0,1]\setminus\{\overline x_a\}$:}
    We now show $v''_a$ exists and is continuous everywhere except possibly at $\overline x_a$.
    Fix $x \in [0,\overline x_a)$.
    As $v''_a$ exists a.e. then take any sequence ${(x_n)}_{n\geq 1}\subseteq [0,\overline x_a)$ such that $x_n\to x$ and $v''_a(x_n)$ exists for every $n$.
    Then, $(v'_a(x_n),v''_a(x_n))\in J^{2,+}_{[0,1]}v_a(x_n)\cap J^{2,-}_{[0,1]}v_a(x_n)$, as this is true if and only if $v_a$ is twice differentiable at $x_n$ \citep[p. 15]{CrandallIshiiLions1992BullAMS}.
    Hence, $(x_n,v_a(x_n),v'_a(x_n))\to (x,v_a(x),v'_a(x))$.
    $\forall y \in [0,\overline x_a)$, $F_a(x,v_a(x),M)\leq 0$ for all $M\geq \overline M:=\max_{x \in [0,\overline x_a]}\sqrt{2c}r\sqrt{v_a(x)-x}$.
    Hence, $(v'_a(x),\overline M)\in J^{2,+}_{[0,1]}v_a(x)$.
    Together with convexity of $v_a$ on $[0,\overline x_a)$, this implies that $v''_a(x_n)\in [0,\overline M]$ for all $n$, and then, by compactness, $v''_a(x_n)$ has a convergent subsequence. 
    Take any convergent subsequence and denote its limit as $v''_\infty$.
    As $F_a$ is continuous, $0=F_a(x_n,v_a(x_n),v''_a(x_n))\to F_a(x,v_a(x),v''_\infty)=0\Longrightarrow (v'_a(x),v''_\infty) \in J^{2,+}_{[0,1]}v_a(x)\cap J^{2,-}_{[0,1]}v_a(x)$, ensuring that $v_a$ is also twice differentiable at $x$, for any $x \in [0,\overline x)$, and $v''_a(x)=v''_\infty$.
    This implies $v''_a$ exists everywhere in $[0,\overline x_a)$.
    Moreover, as $v''_\infty \geq 0$ and $F_a(x,v_a(x),M')<F_a(x,v_a(x),M)$ for any $M'>M\geq 0$, we must then have $v''_\infty$ being the limit of any convergent subsequence of ${(v''_a(x_n))}_{n\geq 1}$, and so, the limit of the original sequence: $v''_a(x_n)\to v''_\infty=v''_a(x)$, and we obtain that $v''_a\in \mathcal C^2$ on $[0,\overline x_a)$.
    A symmetric argument holds for $x \in (\overline x_a, 1]$.
    
    \textbf{Zero derivative at 0:} Suppose $v'_a(0)>0$. Then, $(v'_a(0)/2,2v''_a(0))\in J^{2,-}_{[0,1]}v_a(0)$\footnote{
        To see this, define $f(x)=v_a(0)+\frac{v'_a(0)}{2}x+v''_a(0)x^2$, noting that $f(x)\leq v_a(x)$ for small enough $x$.
    } and\\ $\max\{F_a(0,v_a(0),2v''_a(0)), \, B(0,v'_a(0)/2)\}<0$, contradicting that $v_a$ is supersolution.
    
    \textbf{Necessary conditions for nondifferentiability at $\overline x_a$:} 
    (a) and (b) follow from there being only concave kinks and only if $b(\overline x_a)=0$.
    If $\overline x_a<1$ and $b(1)=0$, then we must have $v_a$ convex on $[0,1]$ and linear on $[\underline x_a,1]$.
    It follows that $v'_{a,-}(\overline x_a)\geq 1$ (no convex kinks) and $v'_{a,-}\leq 1$, which implies $v'_{a,-}(\overline x_a)=1=v'_{a,+}(\overline x_a)$, and the argument from above extends to show that $v_a \in \mathcal C^2([0,1])$.
    Consequently, we obtain (c) by the contrapositive.
\end{proof}

\subsection{Proof of \hyref{prop:decreasing-control}{Proposition} (Decreasing Control)}
\begin{proof}
    As $a(x) = \frac{1}{rc} v_a''(x)_+$, where $v_a$ is the solution to \hyperref[eq:HJB]{\color{black}(RP)} given $b$, it suffices to show $v_a''$ is non-increasing in the convex region of $v_a$ i.e. on $[0,\underline x_a]$, where $\underline x_a$ is as defined in \hyref{prop:convex-concave}{Proposition}.
    Assume by contradiction $\exists x,y \in [0,\underline x_a]: x > y$ and $v''_a(x)>v''_a(y)$. 
    Then, using the fact that $b$ is non-decreasing on this region,
    \begin{align*}
        0  &< \frac{1}{2rc} \bigl( v''_a(x)^2 - v''_a(y)^2 \bigr)
         && = r[v_a(x)-x]-r[v_a(y)-y]-b(x)v''_a(x)+b(y)v''_a(y)
        \\
         &&& \leq 
         r[v_a(x)-x]-r[v_a(y)-y]-\bigl( b(x)-b(y) \bigr)v''_a(x)
         \leq r[v_a(x)-x]-r[v_a(y)-y]
    \end{align*}
    \noindent hence $1<\frac{v_a(x)-v_a(y)}{x-y}=v_a'(z)$ for some $z \in (y,x)$ (mean value theorem), contradicting $0\leq v'_a\leq 1$ (\hyref{thm:best-response-characterization}{Theorem}).
\end{proof}

\subsection{Proof of \hyref{prop:properties-individual-control}{Proposition} (Control is $\mathcal C^1$)}

\hyref{prop:properties-individual-control}{Proposition} follows from \hyref{thm:best-response-characterization}{Theorem} and the next lemma.

 \begin{lemma}\label{lemma:control-c1-idm}
    If $b \equiv 0$ and $\underline{x}_a^0<1$ the optimal control to \hyperref[eq:HJB]{\color{black}(RP)} is $\mathcal C^1([0,1])$. If $\underline{x}_a^0 =1$ it is $\mathcal C^1([0,1))$
\end{lemma}

\begin{proof}
    Let $v_a$ be a viscosity solution to \hyperref[eq:HJB]{\color{black}(RP)} on $\mathcal O=(0,1)$ when $b \equiv 0$ and $a$ the associated optimal control.
    Define $\underline x_a$ as in \hyref{prop:convex-concave}{Proposition}.
    On $[0,\underline x)$,
        $F(x,v_a,v_a'')=0  \,\,\,       \Longleftrightarrow\,\,\, v_a''(x)=r\sqrt{2c}\sqrt{v_a(x)-x}$,
    and $a$ is continuously differentiable (even infinitely so) on this $[0,\underline{x})$. 
    This proves the lemma for $\underline{x}_a=1$. 
    If $\underline{x}_a<1$, then $v_a(x)=x$ on $(\underline{x}_a,1]$ implying $a$ is $\mathcal{C}^1$ on this interval, with $a'(x)=0$, and $\lim_{x\downarrow \underline x_a}a'(x)=0$.
    For the left derivative, noting ${v_a''(x)}^2=2r^2c(v_a(x)-x)$ for any $x \in[0,\underline x_a)$ and $v_a''(x)>0$, differentiate both sides and obtain
    $a'(x)=v_a'''(x)/rc=r (v'_a(x)-1)/v_a''(x),$
    which is continuous as $v$ is $\mathcal C^2$ on $[0,\underline x_a)$ given $b\equiv 0$ (\hyref{prop:C2}{Proposition}). 
    As $v'_a(x)-1<0$ (\hyref{prop:convex-concave}{Proposition}), then $v_a'''(x)<0$, and $v'_a$ is strictly increasing and strictly concave on this interval.
    Hence, $v_a'(\underline x_a)-v_a'(x)\leq v_a''(x)(\underline x_a-x)$, $\forall x< \underline x_a$.
    Thus, 
    $0\leq (v_a'(\underline{x})-v_a'(x))/v_a''(x)\leq \underline x - x.$
    As 
    $(1-v_a'(x))/v_a''(x)\to 0$ for $x\uparrow \underline x_a$, and $a'_-(\underline x_a)=0$, we obtain that $a$ is $\mathcal C^1$ on $[0,1]$.
\end{proof}

\subsection{Proof of \hyref{prop:properties-upper-reflection}{Proposition} (Control is Convex-Concave)}
\begin{proof}
    Let $v_a$ be a viscosity solution to \hyperref[eq:HJB]{\color{black}(RP)} on $\mathcal O=(0,1)$ when $b \equiv 0$, and $a$ the associated optimal control. Recall that $a\propto v''_a$. 
    Denote by $f'_-$ the left-derivative of $f$ and $f^{(n)}$ its $n$-th order derivative.
    From \hyref{prop:convex-concave}{Proposition}, we have that $v'_{a,-}\leq 1$.
    Owing to the regularity of the solution, and we can derive on $[0,\underline x_a)$:
    \begin{align*}
        v''_a(x)&=r\sqrt{2c}\sqrt{v_a(x)-x}\geq 0,\quad
        &v^{(3)}_a(x)&=r\sqrt{c/2}{(v_a(x)-x)}^{-1/2}(v'_a(x)-1)=r^2c\frac{v'_a(x)-1}{v''_a(x)}\leq 0,\\
        v^{(4)}_a(x)&=r^2c - \frac{v^{(3)}_a(x)^2}{v''_a(x)},\quad
        &v^{(5)}_a(x)&=\frac{v^{(3)}_a(x)^3}{v''_a(x)^2}-2\frac{v^{(3)}_a(x)}{v''_a(x)}v^{(4)}_a(x).
    \end{align*}
    As $v'_a(x)<1$ for $x \in [0,\underline x_a)$, $v^{(3)}_a$ is strictly negative on $[0,\underline x_a)$.
    If, for $x \in (0,\underline x_a)$, $v^{(4)}_a(x)=0$, then $v^{(5)}_a(x)=\frac{v^{(3)}_a(x)^3}{v''_a(x)^2}<0$.
    This implies that if, for $\tilde x \in (0,\underline x_a)$, $v^{(4)}_a(\tilde x)=0$, then $v^{(4)}_a(x)\leq 0$ for any $x \in (\tilde x, \underline x_a)$.
    That is, $\exists \tilde x \in [0,\underline x_a)$ such that $v''_a$ is convex on $[0,\tilde x]$ and concave on $[\tilde x,\underline x_a]$.
    
    Suppose $v'_{a,-}({\underline x_a})=1$.
    As, by \hyref{prop:decreasing-control}{Proposition}, $\lim_{x\uparrow \underline x_a}v^{(3)}_a(x)=0$, we have
    \begin{align*}
        \lim_{x\uparrow \underline x_a} v^{(4)}_a(x)&=r^2 c - \lim_{x\uparrow \underline x_a} \frac{v^{(3)}_a(x)^2}{v''_a(x)}
        =r^2 c - r^2 c\lim_{x\uparrow \underline x_a} \frac{{(v'_a(x)-1)}^2}{2r\sqrt{2c}{(v_a(x)-x)}^{3/2}}
        \\
        &=r^2 c - r^2 c\lim_{x\uparrow \underline x_a} \frac{2}{3}\frac{{(v'_a(x)-1)}v''_a(x)}{r\sqrt{2c}{(v_a(x)-x)}^{1/2}(v'_a(x)-1)}
        =r^2 c - r^2 c\lim_{x\uparrow \underline x_a} \frac{2}{3}=\frac{1}{3}r^2 c>0,
    \end{align*}
    where we used l'H\^{o}pital's rule in the before-last line.
    Consequently, $v''_a$ is convex on $[0,\underline x_a]$.
    
    Suppose now that $v'_{a,-}({\underline x_a})<1$.
    Then $v^{(3)}_a(x)\leq r^2c\frac{v'_{a,-}(\underline x_a)-1}{v''_a(0)}<0$ for any $x \in [0,\underline x_a]$.
    As $v''_a$ is strictly positive, decreasing, $v''_a(x)\to 0$ as $x\to \underline x_a$, $v^{(4)}_a(x)<0$ for all $x<\underline x_a$ close enough to $\underline x_a$.
    Hence, $\exists \tilde x \in [0,\underline x_a)$ such that $v''_a$ is convex on $[0,\tilde x]$ and concave on $[\tilde x,\underline x_a]$.
    
    The fact that $v'_{a,-}({\underline x_a}^-)=1$ if $\underline x_a<1$ follows from the $a$ being $\mathcal C^1([0,1])$ when $\underline x_a<1$ (\hyref{lemma:control-c1-idm}{Lemma}).
    Finally, that $a'_-(\underline x_a)=-\infty$ follows from  $v^{(3)}_a(x)=r^2c\frac{v'_a(x)-1}{v''_a(x)}\leq r^2c\frac{v'_{a,-}(\underline x_a)-1}{v''_a(x)}<0$.
    As the denominator goes to zero as $x$ approaches $\underline x_a$, the result obtains. 
\end{proof}

\subsection{Proof of \hyref{prop:equilibrium-properties}{Proposition} (Decoupling Equilibrium Instability)}
\begin{proof}
    Note that, from \hyref{thm:HJBBC-controlproblem}{Theorem}, $v_a(x)+v_b(x)\leq\sup_{\alpha,\beta}r\int_0^\infty\exp(-rt)(X_t+(1-X_t)-c_a\alpha(X_t)^2-c_b\beta(X_t)^2)dt\leq r\int_0^\infty\exp(-rt)dt=1$.
    From \hyref{prop:convex-concave}{Proposition}, 
    as $v_a$ is (strictly) convex whenever $v_a(x)\geq \,(>)\,x \Longleftrightarrow 0\leq x\leq \overline x_a \,(<\underline x_a)$ and strictly concave elsewhere,
    and $v_b$ is (strictly) convex whenever $v_b(x)\geq \,(>)\,1-x \Longleftrightarrow 1\geq x\geq \overline x_b \,(>\underline x_b)$, and strictly concave elsewhere, we have that $\underline x_a = \overline x_a=:\underline x>0$ and $\overline x_a=\underline x_b=:\overline x<1$.
    This implies that $a(x)=0$ on $[\underline x,1]$ and $b(x)=0$ on $[0,\overline x]$.
    As, from \hyref{prop:decreasing-control}{Proposition} $a$ is nonincreasing and $b$ is nondecreasing, and, from a straightforward modification of the proof of \hyref{prop:properties-individual-control}{Proposition},
    $v_a'''<0$ on $[0,\underline x_a)$ and $v_b'''(x)>0$ on $(\underline x_b,1]$, we obtain that the optimal controls $a$ and $b$ are, respectively, strictly decreasing and strictly increasing.
\end{proof}

\subsection{Proof of \hyref{prop:individualcontrol-supersolution}{Proposition} (Inactive Benchmark and Equilibrium)}
\begin{proof}
    From \hyref{prop:equilibrium-properties}{Proposition}, $a^*(x)>0$ if and only if $x\in [0,\underline x)$, and, from \hyref{thm:best-response-characterization}{Theorem}, $v_a$ is concave on $[\underline x,1]$.
    Hence,
    on $x \in [0,\underline x)$,
    $0=r_a(v_a(x)-x)-\frac{1}{r_a c_a}{[v_a''(x)_+]}^2,$
    and on $x \in [\underline x,1]$ except at most at one point at which $v_a$ is not twice differentiable,
    $0=r_a(v_a(x)-x)-b(x)v_a''(x)\geq r_a(v_a(x)-x).$
    As at the (at most one) nondifferentiability point of $v_a$ there is a concave kink (\hyref{thm:best-response-characterization}{Theorem}), one concludes $v_a$ is a viscosity subsolution to the reflected problem in the inactive benchmark.
    As $v_a^0$ is a viscosity solution to the same problem (and thus a supersolution), from \hyref{lemma:comparisonprinciple-controlproblem}{Lemma}, $v_a^0\geq v_a$. 
    The second part of the proposition follows immediately.
    The same holds for player $B$.
\end{proof}

\subsection{Proof of \hyref{lemma:singleton-stable-region-deterrence}{Lemma} (Deterrence Equilibria Singleton Stable Region)}
\begin{proof}
    We prove the lemma by contradiction.
    Let $(a,b)$ be an equilibrium under parameters such that $\underline x_a^0>\underline x_b^0$ and, for the purpose of contradiction, suppose $\underline x<\overline x$.
    Then, $\underline x< \underline x_a^0$ or $\underline x_b^0<\overline x$.
    This is because, from \hyref{prop:individualcontrol-supersolution}{Proposition},
    $\underline x\leq \underline x_a^0$ or $\underline x_b^0\leq\overline x$, and, 
    by assumption, $\underline x_a^0>\underline x_b^0$.
    Suppose $\underline x<\underline x_a^0$ (the proof is symmetric for the case in which $\underline x_b^0<\overline x$).
    From \hyref{thm:best-response-characterization}{Theorem}, only concave kinks are permissible, and then $v'_{a,-}(\underline x)\geq v'_{a,+}(\underline x)=1$.
    Moreover, from \hyref{prop:individualcontrol-supersolution}{Proposition}, the solution to player $A$'s the inactive benchmark problem, $v_a^0$, is weakly greater than the player's equilibrium value function, $v_a\leq v_a^0$.
    From \hyref{prop:equilibrium-properties}{Proposition}, at an equilibrium, $b(x)=0$ on $[0,\overline x]\supseteq [0,\min\{\overline x,\underline x_a^0\}]$.
    As $v^{0\prime\prime}_a(x)=r_a \sqrt{2c_a }\sqrt{v^0_a(x)-x}\geq r_a \sqrt{2c_a }\sqrt{v_a(x)-x}=v''_a(x)$
    on $[0,\min\{\overline x,\underline x_a^0\}]$
    and as $v'_a(0)=v^{0\prime}_a(0)=0$ (\hyref{prop:C2}{Proposition}), then
    $v'_{a,-}(\underline x)\leq v^{0\prime}_{a,-}(\underline x)< v^{0\prime}_{a,-}(\underline x_a)\leq 1 =v'_{a,+}(\underline x)$, a contradiction.
\end{proof}

\subsection{Proof of \hyref{thm:eqm-characterization}{Theorem} (Characterization of Deterrence Equilibria)}

The proof of the first part of \hyref{thm:eqm-characterization}{Theorem} (characterization of accommodating equilibria) is detailed in the main text; here we prove the second part (characterization of deterrence equilibria).

Let $v_a^{b}$ be the unique viscosity solution to \hyperref[eq:HJB]{\color{black}(RP)} on $\mathcal O=(0,1)$ given $b\in \mathcal C^0([0,1])$ and $x_a^b:=\sup\{x \in [0,1]\,\mid\,v_a^{b}(x)>x\}$, and analogously define $v_b^{a}$ and $x_b^a$ for player $B$, given $a \in \mathcal C^0([0,1])$.
It is straightforward to check that, for $(a^*,b^*)$ such that $x_b^0\leq x_a^{b^*}= x_b^{a^*}\leq x_a^0$, the equilibrium strategies must be given as described in the statement of \hyref{thm:eqm-characterization}{Theorem}.
We then focus on showing that for any $\overline x \in [x_b^0,x_a^0]$, there is a unique strategy profile $(a^*,b^*)$ such that $x_a^{b^*}= x_b^{a^*}=\overline x$.
The proof of \hyref{thm:eqm-characterization}{Theorem} for $\overline x \in (\underline x_b^0,\underline x_a^0)$ follows from the next two lemmata:
\begin{lemma}\label{lemma:deterrence:1}
    For $\overline x\in (0,\underline x_a^0)$, let $v_a$ denote the unique viscosity solution to \hyperref[eq:HJB]{\color{black}(RP)} on $\mathcal O=(0,\overline x)$ when $b \equiv 0$.
    Then, (i) $v_a \in \mathcal C^5([0,\overline x))$, (ii) $v_a$ is convex, (iii) $v'_a$ is concave, (iv) $\exists \tilde x \in [0,\overline x)$ such that $v''_a$ is convex on $[0,\tilde x]$ and concave on $[\tilde x, \overline x)$, and (v) $v'''_a(x)\to -\infty$ as $x \uparrow \overline x$.
\end{lemma}
\begin{proof}
    That there is a unique viscosity solution to \hyperref[eq:HJB]{\color{black}(RP)} on $\mathcal O=(0,\overline x)$ when $b \equiv 0$ follows from a straightforward modification of \hyref{thm:HJBBC-controlproblem}{Theorem}.
    Properties (i)-(v) follow from adjusting the proofs of \hyref{prop:properties-individual-control}{Propositions} and \ref{prop:properties-upper-reflection}.
\end{proof}
\begin{lemma}\label{lemma:deterrence:2}
    Let $\overline x \in (0,1)$ and fix $b\in \mathcal C^0([0,1])$ such that (i) $b(x)=0$ for $x\leq \overline x$, (ii) $b'(x)>0$ on $(\overline x,1]$, (iii) $\lim_{x\downarrow \overline x}b'(x)= \infty$.
    Then, $v_a^b(x)\leq x$ for $x \geq \overline x$.
\end{lemma}
\begin{proof}
    Suppose not. Then, $v_a^b(\overline x)>\overline x \Longrightarrow v_a^{b\prime\prime}(\overline x)>0$, and, by \hyref{prop:properties-individual-control}{Proposition}, $v_a^{b}$ is $\mathcal C^3$ locally at $\overline x$ with $v'''_a(x)<0$ in a neighborhood of $\overline x$.
    Then, as $b(\overline x)=0$, for small $\epsilon>0$, 
    $
    F_a(\overline x,v_a^b(\overline x), v_a^{b\prime\prime}(\overline x))=F_a(\overline x+\epsilon,v_a^b(\overline x+\epsilon), v_a^{b\prime\prime}(\overline x+\epsilon))=0
    \Longleftrightarrow 
    0=(F_a(\overline x,v_a^b(\overline x), v_a^{b\prime\prime}(\overline x))-F_a(\overline x+\epsilon,v_a^b(\overline x+\epsilon), v_a^{b\prime\prime}(\overline x+\epsilon)))/\epsilon=
    r[(v_a^b(\overline x+\epsilon)-v_a^b(\overline x))/\epsilon-1]-\frac{1}{2rc}(v_a^{b\prime\prime}(\overline x+\epsilon)^2-v_a^{b\prime\prime}(\overline x)^2/\epsilon-v_a^{b\prime\prime}(\overline x+\epsilon)b(\overline x+\epsilon)/\epsilon $.
    Given that $\lim_{\epsilon\downarrow 0}|r[(v_a^b(\overline x+\epsilon)-v_a^b(\overline x))/\epsilon-1]-\frac{1}{2rc}(v_a^{b\prime\prime}(\overline x+\epsilon)^2-v_a^{b\prime\prime}(\overline x)^2)/\epsilon|= |r(v_a^{b\prime}(\overline x)-1)-\frac{1}{rc}v_a^{b\prime\prime}(\overline x)v_a^{b\prime\prime\prime}(\overline x)|<\infty$ due to $v_a^b$ being locally $\mathcal C^3$, and given that $b(\overline x+\epsilon)/\epsilon=(b(\overline x+\epsilon)-b(\overline x))/\epsilon \to \infty$ as $\epsilon \downarrow 0$, by continuity, $\exists \bar \epsilon: \forall \epsilon\in (0,\bar \epsilon)$, 
    $v_a^{b\prime\prime}(\overline x+\bar \epsilon)b(\overline x+\epsilon)/\epsilon >2[r|v_a^{b\prime}(\overline x)-1|+\frac{1}{rc}v_a^{b\prime\prime}(\overline x)|v_a^{b\prime\prime\prime}(\overline x)|]>r[(v_a^b(\overline x+\epsilon)-v_a^b(\overline x))/\epsilon-1]-\frac{1}{2rc}(v_a^{b\prime\prime}(\overline x+\epsilon)^2-v_a^{b\prime\prime}(\overline x)^2/\epsilon$, a contradiction to $0=(F_a(\overline x,v_a^b(\overline x), v_a^{b\prime\prime}(\overline x))-F_a(\overline x+\epsilon,v_a^b(\overline x+\epsilon), v_a^{b\prime\prime}(\overline x+\epsilon)))/\epsilon$.
\end{proof}
We now take care of showing that $\overline x \in \{\underline x_a^0,\underline x_b^0\}$ also pins-down an equilibrium as described.

If $a\equiv 0$ on $[x_b^0,1]$, then $v_b^{a}=v_b^0$ on $[x_b^0,1]$ and so $b^0$ is a best response to $a$.
We then need that, if $a$ is a best response to $b^0$, then $x_a^{b^0}=x_b^0$.
We prove this in two steps.

First, we show $x_a^{b^0}\geq x_b^0$.
    Suppose not. 
    Then $v_a^{b^0}(x)=x$ on $[x_a^{b^0},x_b^0]$ and, by \hyref{thm:best-response-characterization}{Theorem}, $v_a^{b^0} \in \mathcal C^2([0,1])$.
    Let $w_a(x):=\mathbf 1_{x<x_a^{b^0}}v_a^{b^0}(x)+\mathbf 1_{x\geq x_a^{b^0}} x$.
    As $v_a^{b^0}\in \mathcal C^2([0,1])$ is a viscosity solution to \hyperref[eq:HJB]{\color{black}(RP)} on $\mathcal O=(0,1)$ given $b=b^0$, and as $x_a^{b^0}<x_b^0$, it is straightforward to verify $w_a$ is a viscosity solution to \hyperref[eq:HJB]{\color{black}(RP)} on $\mathcal O=(0,1)$ given $b\equiv 0$.
    However, $x_a^{b^0}<x_b^0<x_a^0$, which contradicts uniqueness of the viscosity solution to the latter problem (\hyref{thm:HJBBC-controlproblem}{Theorem}).
Second, we show that $x_a^{b^0}\leq x_b^0$. 
    Suppose not. 
    Take any $\overline x \in (x_b^0,x_a^{b^0})$ and let $(a^*,b^*)$ be the unique equilibrium such that $a^*(\overline x)=b^*(\overline x)=0$ (\hyref{lemma:deterrence:1}{Lemmata} and \ref{lemma:deterrence:2}).
    
    \textbf{Claim 1: $v_a^{b^0}\geq v_a^{b^*}$ on $[0,x_a^{b^0}]$ and $v_a^{b^0}> v_a^{b^*}$ on $(\overline x,x_a^{b^0}]$.} 
    Let $\underline w_a(x):=\mathbf 1_{x\leq \overline x}v_a^{b^*}(x)+\mathbf 1_{x>\overline x} \overline x$.
    Since (i) $b^0\geq b^*$ (\hyref{prop:individualcontrol-supersolution}{Proposition}) and $b^0>0\equiv b^*$ on $(x_b^0,\overline x)$, and (ii) $v_a^{b^0}$ and $v_a^{b^*}$ are strictly convex on $[0,\overline x)$, then $\underline w_a$ is a subsolution to \hyperref[eq:HJB]{\color{black}(RP)} on $\mathcal O=(0,1)$ given $b=b^0$, and, in particular, $\underline w_a\leq v_a^{b^0}$ (\hyref{lemma:comparisonprinciple-controlproblem}{Lemma}).
    Hence, $v_a^{b^0}\geq  v_a^{b^*}$ on $[0,\overline x]$.
    And, on $(\overline x,x_a^{b^0}]$, by definition of these thresholds, $v_a^{b^0}(x)\geq x>v_a^{b^*}$.
    
    \textbf{Claim 2: $v_a^{b^0}\leq v_a^{b^*}$ on $[x_a^{b^0},1]$.} 
    Let $\overline w_a(x):=\mathbf 1_{x\geq x_a^{b^0}}v_a^{b^0}(x)+\mathbf 1_{x<x_a^{b^0}} x$.
    Since (i) $b^0\geq b^*$ (\hyref{prop:individualcontrol-supersolution}{Proposition}), and (ii) $v_a^{b^0}$ and $v_a^{b^*}$ are strictly concave on $(x_a^{b^0},1]$, then $\overline w_a$ is a subsolution to \hyperref[eq:HJB]{\color{black}(RP)} on $\mathcal O=(0,1)$ given $b=b^*$, and, in particular, $\overline w_a\leq v_a^{b^*}$.
    Hence, $v_a^{b^0}\leq  v_a^{b^*}$ on $[x_a^{b^0},1]$.
    
    However, claims 1 and 2 clearly entail a contradiction: $v_a^{b^*}(x_a^{b^0})<v_a^{b^0}(x_a^{b^0})\leq v_a^{b^*}(x_a^{b^0})$.

\end{document}